\documentclass[twocolumn]{aastex62}
\usepackage{natbib}

\newcommand{\angstrom}{\textup{\angstrom}}

\graphicspath{{./}{figures_submit/}}

\accepted{2020 June 29}

\submitjournal{ApJ}

\shorttitle{BLR Metallicity of a $z=7.54$ Quasar}
\shortauthors{Onoue et al.}

\begin{document}

\title{No Redshift Evolution in the Broad Line Region Metallicity up to $z=7.54$: Deep NIR Spectroscopy of ULAS J1342+0928}

\correspondingauthor{Masafusa Onoue}
\email{onoue@mpia-hd.mpg.de}

\author[0000-0003-2984-6803]{Masafusa Onoue}
\affiliation{Max-Planck-Institut f\"ur Astronomie, K\"onigstuhl 17, D-69117 Heidelberg, Germany}

  \author[0000-0002-2931-7824]{Eduardo Ba\~nados}
  \affiliation{Max-Planck-Institut f\"ur Astronomie, K\"onigstuhl 17, D-69117 Heidelberg, Germany}
  \author[0000-0002-5941-5214]{Chiara Mazzucchelli}
  \affiliation{European Southern Observatory, Alonso de C\'ordova 3107, Vitacura, Regi\'on Metropolitana, Chile}
   
  \author[0000-0001-9024-8322]{Bram P. Venemans}
  \affiliation{Max-Planck-Institut f\"ur Astronomie, K\"onigstuhl 17, D-69117 Heidelberg, Germany}

  \author[0000-0002-4544-8242]{Jan-Torge Schindler}
  \affiliation{Max-Planck-Institut f\"ur Astronomie, K\"onigstuhl 17, D-69117 Heidelberg, Germany}

  \author[0000-0003-4793-7880]{Fabian Walter}
  \affiliation{Max-Planck-Institut f\"ur Astronomie, K\"onigstuhl 17, D-69117 Heidelberg, Germany}

  \author[0000-0002-7054-4332]{Joseph F. Hennawi}
  \affiliation{Department of Physics, Broida Hall, University of California at Santa Barbara, Santa Barbara, CA 93106, USA}

  \author[0000-0001-6102-9526]{Irham Taufik Andika}
  \affiliation{Max-Planck-Institut f\"ur Astronomie, K\"onigstuhl 17, D-69117 Heidelberg, Germany}

  \author[0000-0003-0821-3644]{Frederick B. Davies}
  \affiliation{Lawrence Berkeley National Laboratory, 1 Cyclotron Rd, Berkeley, CA 94720, USA}

 \author[0000-0002-2662-8803]{Roberto Decarli}
 \affiliation{INAF--Osservatorio di Astrofisica e Scienza dello Spazio di Bologna, via Gobetti 93/3, I-40129, Bologna, Italy}

  \author[0000-0002-6822-2254]{Emanuele P. Farina}
  \affiliation{Max-Planck-Institut f\"ur Astronomie, K\"onigstuhl 17, D-69117 Heidelberg, Germany}
  \affiliation{Max-Planck-Institut f\"ur Astrophysik, Karl-Schwarzschild-Stra{\ss}e 1, D-85748, Garching bei M\"unchen, Germany}

 \author[0000-0003-3804-2137]{Knud Jahnke}
  \affiliation{Max-Planck-Institut f\"ur Astronomie, K\"onigstuhl 17, D-69117 Heidelberg, Germany}

 \author[0000-0002-7402-5441]{Tohru Nagao}
  \affiliation{Research Center for Space and Cosmic Evolution, Ehime University, Matsuyama, Ehime 790-8577, Japan.}

  \author[0000-0001-8537-3153]{Nozomu Tominaga}
  \affiliation{Department of Physics, Faculty of Science and Engineering, Konan University, 8-9-1 Okamoto, Kobe, Hyogo 658-8501, Japan}
  \affiliation{Kavli Institute for the Physics and Mathematics of the Universe (Kavli IPMU, WPI), The University of Tokyo, 5-1-5 Kashiwanoha, Kashiwa, Chiba 277-8583, Japan}

\author[0000-0002-7633-431X]{Feige Wang}
\altaffiliation{NHFP Hubble Fellow}
\affil{Steward Observatory, University of Arizona, 933 North Cherry Avenue, Tucson, AZ 85721, USA}
\affil{Department of Physics, University of California, Santa Barbara, CA 93106-9530, USA}



\begin{abstract}
We present deep ($9$ hours) Gemini-N/GNIRS near-infrared spectroscopic observations of ULAS J1342+0928, a luminous quasar at $z=7.54$.
Various broad emission lines were detected, as well as the underlying continuum and iron forests over the rest-frame wavelength 970--2930\AA.
There is a clear trend that higher-ionization emission lines show larger blueshifts
with C{\sc iv} $\lambda1549$ exhibiting $5510^{+240}_{-110}$ km s$^{-1}$ blueshift 
with respect to the systematic redshift from the far-infrared [C{\sc ii}] $158\micron$ emission line.
Those high ionization lines have wide profiles with FWHM more than 10000 km s$^{-1}$.
A modest blueshift of $340^{+110}_{-80}$ km s$^{-1}$ is also seen in Mg{\sc ii}, the lowest ionization line identified in the spectrum.
The updated Mg{\sc ii}-based black hole mass of $M_\mathrm{BH}=9.1_{-1.3}^{+1.4} \times 10^8 M_\odot$ and the Eddington ratio of   $L_\mathrm{bol}/L_\mathrm{Edd}=1.1_{-0.2}^{+0.2}$
confirm that ULAS J1342+0928 is powered by a massive and actively accreting black hole.
There is no significant difference in the emission line ratios such as Si{\sc iv}/C{\sc iv} and Al{\sc iii}/C{\sc iv} when compared to lower-redshift quasars in a similar luminosity range, which suggests early metal pollution of the broad-line-region clouds.
This trend also holds for the Fe{\sc ii}/Mg{\sc ii} line ratio, known as a cosmic clock that traces the iron enrichment in the early universe.
Different iron templates and continuum fitting ranges were used 
to explore how the Fe{\sc ii}/Mg{\sc ii} measurement changes as a function of spectral modeling.
Quasars at even higher redshift or at fainter luminosity range ($L_\mathrm{bol}\lesssim10^{46}$ erg s$^{-1}$) are needed to probe the sites of early metal enrichment and a corresponding change in the Fe{\sc ii}/Mg{\sc ii} ratio.

\end{abstract}

\keywords{dark ages, reionization -- quasars:general -- quasars: supermassive black holes -- quasars: individual (ULAS J134208.10+092838.35)}

\section{Introduction} \label{sec:Sec1}
Quasars are among the most luminous objects in the universe.
Powered by mass accretion onto the central supermassive black holes (SMBHs),
quasars can be observed out to the epoch of cosmic reionization (at redshift $z\gtrsim6$), in which ultraviolet (UV) photons from the first generation objects light up the dark universe  \citep[e.g.,][]{Fan06, Planck16}.
Observations have reached this crucial epoch, e.g., by finding more than $200$ quasars at $z>5.7$ \citep[e.g.,][]{Fan01, Willott10b, Venemans13, Reed15, Banados16, Jiang16, Matsuoka16, Wang18}.
Six quasars have been identified at $z>7.0$ to date \citep{Mortlock11, Banados18, Wang18, Matsuoka19a, Matsuoka19b, Yang19}.
A surprising fact drawn from those luminous quasars is that SMBHs are in place with $M_\mathrm{BH}\sim10^{9-10}M_\odot$ in the early universe, which are as massive as the most massive SMBHs known in the entire cosmic history \citep[e.g.,][]{Wu15}.
This puts a stringent constraint on the formation scenario of SMBHs.
The possible pathways to form such gigantic SMBHs at $z\gtrsim6$ are either through remnants BHs of population-{\sc III} (Pop-III) stars experiencing
super-Eddington phases during their initial growth, formation of gigantic seed black holes ($10^{5-6} M_\odot$) through direct collapses of chemically pristine gas clouds, or runaway stellar collisions within dense metal-poor clusters, which leave $10^{3-4} M_\odot$ seed BHs \citep[e.g.,][]{Greene20, Inayoshi20}.

Chemical enrichment is another independent diagnostic to explore the early universe with high-redshift quasars \citep{Hamann99, Maiolino19}.
Unobscured quasars generally show broad emission lines in their rest-frame UV spectra with their typical line widths of several $1000$ km s$^{-1}$.
These lines originate from the broad-line-region (BLR) gas, which orbits the central black holes at sub-parsec scale \citep[e.g.,][]{GRAVITY18}.
The BLR gas-phase metallicity can be inferred from the observed emission line ratios through photoionization modeling.
Commonly used line combinations that are most sensitive to the BLR metallicity are, for example, N{\sc v}$\lambda1240$/He{\sc ii}$\lambda1640$ and N{\sc v}$\lambda1240$/C{\sc iv}$\lambda1549$.
The BLR metallicity is thought to represent gas that was subject to the star-formation history of the host galaxies, making quasars unique probes of the chemical enrichment.

Although quasar surveys now probe the first billion years of the universe,
there has been no clear evidence of redshift evolution in the BLR metallicity up to $z\sim7$.
The BLR clouds generally show super solar metallicity ($Z\sim5Z_\odot$; e.g., \citealt{Hamann92, Hamann93, Dietrich03a, Nagao06b}).
\citet{Nagao06b} show composite spectra of $2.0\leq z\leq4.5$ quasars, showing almost constant BLR line ratios at fixed luminosity over the wide redshift range.
Similar measurements have also been done for $z\gtrsim6$ quasars \citep[e.g.,][]{Jiang07, Juarez09, DeRosa11, DeRosa14, Mazzucchelli17, Tang19}, only to find possibly ubiquitous emission line properties of unobscured quasars.
Those studies suggest a rapid and intense chemical enrichment at the centers of the host galaxies, where star formation is most active.
Narrow line regions (NLR) spreading over host-galaxy scale also trace the host metallicity.
The NLR gas metallicity is less extreme than that of the BLR, yet still close to or above the solar value \citep{StorchiBergmann90}, with no significant redshift evolution up to $z\sim5$ \citep{Nagao06a, MatsuokaK09, MatsuokaK11}.

Among various BLR emission lines, the line ratio of UV Fe{\sc ii} complexes and Mg{\sc ii}$\lambda2798$ is of particular interest,
since this quantity serves as a ``cosmic clock"  \citep{Hamann93, Yoshii98}.
Supernovae nucleosynthesis predicts that iron enrichment is delayed from $\alpha$-element by $\sim1 \mathrm{Gyr}$ due to the longer timescale of type-Ia supernovae (SNe Ia) than that of core-collapse type-II supernovae (SNe II) \citep[e.g.,][]{Greggio83, Matteucci86}.
Attempts have been made to detect the delayed iron enrichment in the early universe using Fe{\sc ii}/Mg{\sc ii}, as the two ions have similar ionizing potentials and their wavelengths overlap with each other.
However, no significant redshift evolution has been identified yet up to $z\sim7$, albeit the uncertainties are large in many cases \citep[e.g.,][]{Kawara96, Yoshii98, Thompson99, Iwamuro02, Iwamuro04, Dietrich02, Dietrich03,Barth03, Freudling03,Maiolino03,  Jiang07, Kurk07, DeRosa11, DeRosa14, Mazzucchelli17, Shin19}.

This present paper shows our measurements of rest-frame UV BLR emission lines of a $z=7.54$ quasar, ULAS J1342+0928 \citep{Banados18}, using a deep  NIR spectrum ($9$ hours on source) taken with the Gemini Near-InfraRed Spectrograph (GNIRS) at the Gemini North telescope.
From the first NIR dataset obtained with Gemini/GNIRS and  Folded-port InfraRed Echellette (FIRE) at the Magellan Baade telescope, ULAS J1342+0928 exhibits broad emission lines such as Ly$\alpha$, C{\sc iii]}, C{\sc iv}, and Mg{\sc ii} \citep{Banados18}.
From a simple modeling of quasar continuum and the Mg{\sc ii} line profile, the Mg{\sc ii}-based virial BH mass was estimated to be\footnote{The given SMBH mass and Eddington ratio reported in \citet{Banados18} are modified to the cosmology in this paper, because \citet{Banados18} use the Planck cosmology with $H_0=67.7$ km s$^{-1}$ Mpc$^{-1}$, $\Omega_m=0.307$, and $\Omega_\Lambda=0.693$ \citep{Planck16}.}  $M_\mathrm{BH}=7.6_{-1.9}^{+3.2}\times10^8M_\odot$ with its Eddington ratio of $L_\mathrm{bol}/L_\mathrm{Edd}=1.5_{-0.4}^{+0.5}$.
The main aim of this paper is a detailed spectral modeling of ULAS J1342+0928, in which the iron pseudo-continuum that spreads over $\lambda_\mathrm{rest}\approx2000$--$3000$\AA\ is taken into account.
The structure of this paper is as follows:
Our observations and data reduction are described in Section~\ref{sec:data}.
The spectral analysis of the obtained NIR spectrum is presented in Section~\ref{sec:spec_model}.
The continuum and emission line properties, as well as the measurements of line flux ratios and SMBH mass are presented in Section~\ref{sec:spec_result}.
The systematic uncertainties associated with the Mg{\sc ii} and Fe{\sc ii} measurements, and the origin of early chemical enrichment in the early universe are discussed in Section~\ref{sec:discussion}.
The summary and future prospects are given in Section~\ref{sec:summmary}.

Throughout this paper, the magnitudes quoted are in the AB system.
We adopt a standard $\Lambda$CDM cosmology with $H_0=70$ km s$^{-1}$ Mpc$^{-1}$, $\Omega_m=0.3$, and $\Omega_\Lambda=0.7$.
The [C{\sc ii}] 158$\micron$ redshift presented in \citet[][$z_\mathrm{[CII]}=7.5400\pm0.0003$]{Banados19} is used as the systemic redshift of ULAS J1342+0928.
The age of the universe was $680$ Myr at this redshift in the chosen cosmology.

\section{Data} \label{sec:data}
\subsection{Gemini/GNIRS Spectroscopy} \label{sec:obs}
The spectrum presented in this paper was taken in two different Gemini/GNIRS observation runs.
The first run was on March 31 and April 3, 2017 with a 4.7 hours integration  (GN-2017A-DD-4; PI: E.Ba\~nados).
The data was presented in \citet{Banados18}.
The second observing run was executed between May 27 and June 2, 2019 (GN-2019A-FT-115; PI: M.Onoue), with a 4.3 hours integration.
In this paper, both GNIRS datasets of ULAS J1342+0928 were combined, which results in a total integration time of $9.0$ hours.

Both runs were executed with the same setup.
With the 31.7 l/mm grating and the short camera (0.15 arcsec per pixel), the cross-dispersed mode was used to cover the observed wavelengths of $\lambda_\mathrm{obs}\sim0.9$--$2.5\micron$, which corresponds to the rest-frame wavelengths of $\lambda_\mathrm{rest}=$ 970--2930\AA.
The slit width was chosen to be $0\farcs675$ to perform our spectroscopy with a spectral resolution of $R\sim760$.
The single exposure time was set to 300 sec with the standard ABBA nodding offsets between exposures.
The observation was carried out in good weather conditions with the seeing size of $\sim 0\farcs4$--$0\farcs9$ and at airmass $\sim 1.1$--$1.5$ (with a few exposures at $<1.8$).

\subsection{Data Reduction} \label{sec:pypeit}
Data reduction was performed with {\tt PypeIt}, an open source spectroscopic data reduction pipeline developed by
\citet[][]{Pypeit20}.
Since the target was observed on six different dates, the raw data taken on the same date were processed with the pipeline to obtain calibrated one-dimensional (1D) spectra. 
Each exposure was bias-subtracted and flat-fielded using standard procedures.
The wavelength solution was obtained by comparing the spectrum of the sky with the prominent OH \citep[][]{Rousselot00} and water lines\footnote{\href{https://hitran.org/}{https://hitran.org/}}.
After removing contamination from cosmic ray using the algorithm of \citet{vanDokkum01}, the pipeline optimally subtracts the background by modeling the sky emission with a b-spline function which follows the curvature of the spectrum on the detector \citep[][]{Kelson03}.
The 1D spectrum of the quasar was extracted for each exposure using optimal weighting.

The individual 1D spectra were flux calibrated by fitting for the the order-by-order sensitivity function derived from the A-type stars observed before or after the target exposures.
The initially fluxed 1D spectra from each night were then co-added across the orders to compute a single co-added and fluxed spectrum, but with the telluric absorption still present.
Then a telluric model was directly fitted to the co-added quasar spectrum using the telluric model grids produced from the Line-By-Line Radiative Transfer Model \citep[{\tt LBLRTM} \footnote{\url{http://rtweb.aer.com/lblrtm.html}};][]{Clough05}. 
Specifically, a principal component analysis (PCA) model of the quasar continuum was constructed spanning from 1216\AA\ to 3100\AA\ in the quasar rest-frame following the procedure described in \citet{Davies18}. 
This models the quasar as a linear combination of a mean quasar spectrum and seven PCA components. 
A joint fit was performed for the telluric absorption and PCA continuum
to obtain the 7 PCA coefficients, and overall normalization of the spectrum,  as well as the resolution,  an overall spectral shift, and four parameters describing the atmosphere (airmass, pressure, temperature, and water vapor). 

\begin{figure*}[hp!]
\centering
 \includegraphics[width=0.95\linewidth]{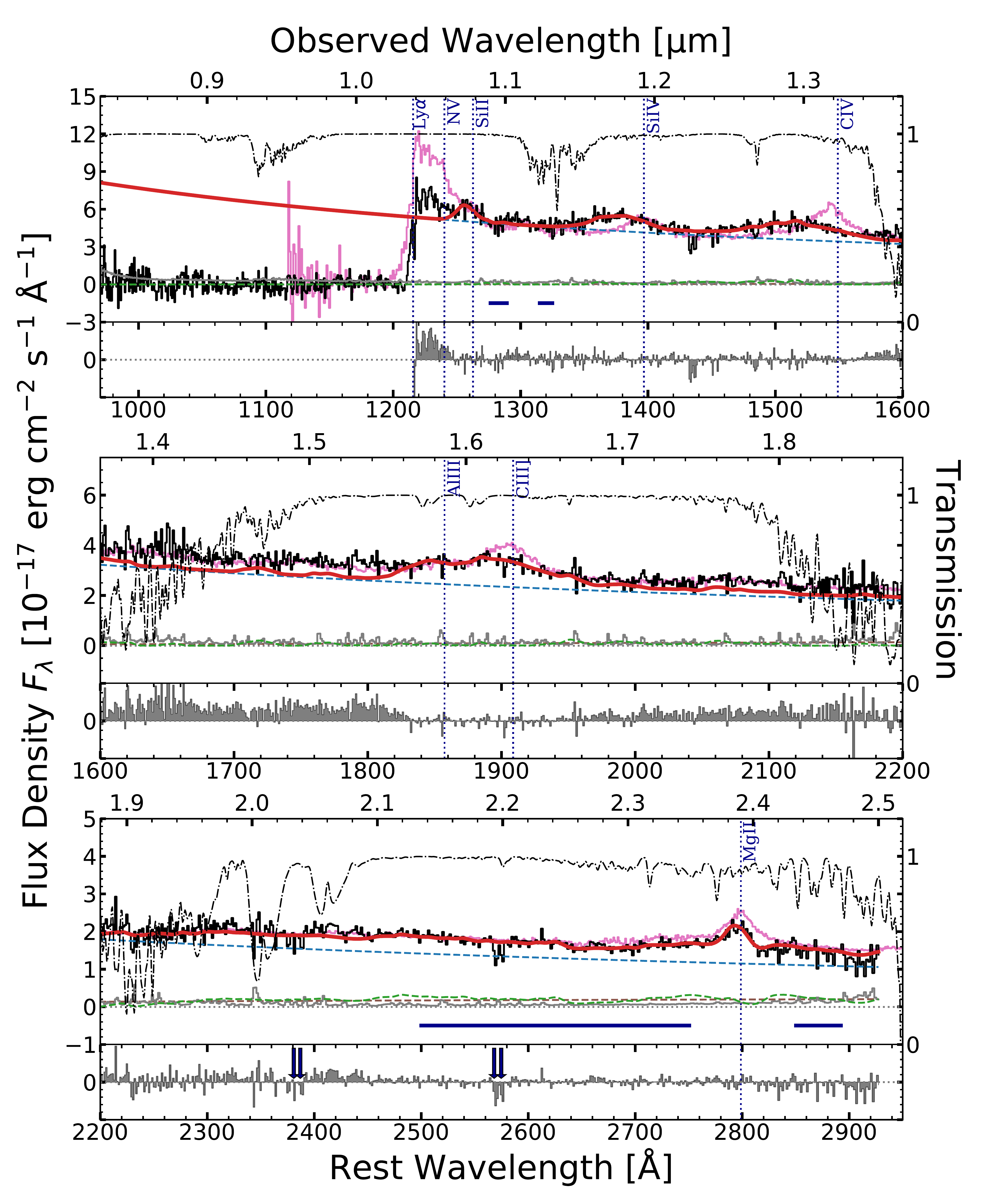}
\caption{
 Gemini/GNIRS spectrum of ULAS J1342+0928 at $\lambda_\mathrm{rest}=$ 970--2930 \AA\ (black) with its error spectrum (grey).
 The spectrum is shifted to rest frame with the [C{\sc ii}] redshift measured in \citet{Banados19}.
Both the rest-frame and the observed wavelength are indicated at the bottom and top of each panel, respectively.
The dashed lines show the best-fit decomposition of power-law continuum (blue), Balmer continuum (brown), and the scaled iron template from \citet[][green]{Tsuzuki06}. 
The red line shows the sum of those three components and emission lines fitted with single Gaussian profiles.
In each of the three wavelength regions, the residual flux is shown in the bottom subpanel.
The composite spectrum of $z\sim6$ quasars from \citet{Shen19} is overplotted (magenta), normalized by the $J$, $H$, and $K_\mathrm{s}$-band magnitudes reported in \citet{Banados18} from top to bottom panel.
The wavelength ranges used for continuum fitting are indicated with solid blue lines.
The iron templates were fitted in the reddest two windows.
The black dot-dash line shows the atmospheric transmission curve obtained for the spectrum taken on one of the observing dates (April 3, 2017).
Two $z>6$ Mg{\sc ii} doublet absorbers reported in \citet{Cooper19} are indicated with arrows.
} \label{fig:spec_all}
\end{figure*}

Those six initially co-added spectra were then resampled in the velocity space by measuring the inverse-square weighted average at $200$ km s$^{-1}$ bins.
Absolute flux calibration was performed by scaling the spectra to the $J$-band photometry reported in \citet[][$J=20.30\pm0.02$]{Banados18}.
This procedure effectively corrects for any slit loss.
Finally, all six fluxed and telluric-corrected 1D spectra were co-added to obtain the final spectrum.
Figure~\ref{fig:spec_all} shows this final GNIRS spectrum in rest frame, in which
the atmospheric transmission curve as a function of the observed wavelength is also shown.

\section{Spectral Analysis} \label{sec:spec_model}
A multi-component (continuum+iron+emission lines) fit was applied to the GNIRS spectrum of ULAS J1342+0928.
Three components were considered to decompose the observed continuum: power-law continuum ($F_\lambda^\mathrm{PL}$), Balmer continuum  ($F_\lambda^\mathrm{BC}$), and iron pseudo-continuum  ($F_\lambda^\mathrm{Fe{\sc II}+Fe{\sc III}}$).
The free parameters are the scale factor and slope of the power-law continuum, and the scale factor of iron pseudo-continuum.
The Balmer continuum is tied to the power-law continuum with a fixed relative scale, as described in detail below.

\subsection{Continuum}  \label{sec:spec_model_cont}
The quasar continuum needs to be subtracted from the spectrum to measure the profiles of broad emission lines.
A sum of two continuum components were considered: power-law ($F_\lambda^\mathrm{PL}=F_{0} \lambda^{\alpha_\lambda}$) and the Balmer continuum.
For the latter, \citet{Grandi82} introduced the following formula:
\begin{equation}
F_\lambda^\mathrm{BC}= F_\mathrm{0}^\mathrm{BE} B_\lambda(T_e) (1-\mathrm{exp}(-\tau_\mathrm{BE}(\lambda/\lambda_\mathrm{BE})^3)),  \label{eq:Balmer}
\end{equation}
where $B_\lambda(T_e)$ is the Planck function at electron temperature $T_e$, and  $\tau_\mathrm{BE}$ is optical depth at the Balmer edge $\lambda_\mathrm{BE} (=3646 \mathrm{\AA})$.
The strength of the Balmer continuum flux cannot be well constrained for the obtained spectrum due to the limited wavelength coverage and the degeneracy with the power-law and iron components.
In this study, the normalization factor $F_\mathrm{0}^\mathrm{BE}$ was fixed to be $30$\% of the power-law continuum at $\lambda_\mathrm{rest}=3675$\AA.
The two parameters, $T_e$ and  $\tau_\mathrm{BE}$, were fixed to be $T_e=15,000$K and  $\tau_\mathrm{BE}=1$, respectively.
Those assumptions were also made in the literature \citep{Dietrich03, Kurk07, DeRosa11, Mazzucchelli17, Shin19}.

Different normalization of the Balmer continuum were also tested from $0$ to $100$\% of the power-law continuum at $\lambda_\mathrm{rest}=3675$\AA\ in order to address the validness of our assumption and the potential systematic uncertainties on the Fe{\sc ii} and Mg{\sc ii} flux measurements.
As a result, the normalization factor of $\sim0$--$30$\%\ gave reasonable spectral decomposition with little changes in the Fe{\sc ii}/Mg{\sc ii} flux ratios (within the $1\sigma$ uncertainty of our fiducial value).
Therefore, our assumption of the normalization factor is robust to the extent of giving accurate flux measurements of Fe{\sc ii} and Mg{\sc ii}.
More details are described in Section~\ref{sec:discussion_continuum}.

\subsection{Iron Emission Lines} \label{sec:spec_model_iron}
Two empirical templates were used to fit the UV iron (Fe{\sc ii}+Fe{\sc iii}) pseudo-continuum for comparison.
For the UV Fe{\sc ii} measurements, 
one of the most frequently used iron templates for the UV Fe{\sc ii} flux measurements is from \citet[][hereafter VW01 template]{Vestergaard01}, while another template from \citet[][hereafter T06 template]{Tsuzuki06} has also been used recently   \citep[e.g.,][]{Sameshima17, Woo18, Shin19}.
Both templates are from a high resolution spectrum of {\sc I} Zw {\sc I}, a narrow-line Seyfert 1 galaxy at $z=0.061$, which exhibits narrow and strong iron emission lines.
The main difference between these two templates is their behaviour around the Mg{\sc ii} line.
The T06 template takes into account the Fe{\sc ii} contribution underneath the Mg{\sc ii} line, using their model spectrum based on a photoionization calculation of the BLR clouds.
On the other hand, the VW01 template has no Fe{\sc ii} flux in this region, which in turn overestimates the actual Mg{\sc ii} flux in the line fitting.
For this reason, the T06 template was selected as our primary template to accurately measure the Fe{\sc ii} and Mg{\sc ii} fluxes.
The T06 template was augmented with the VW01 template at $\lambda_\mathrm{rest}\leq2200$\AA.
Following an empirical correction for the over-subtracted Fe{\sc ii} flux of the VW01 template, constant flux was added to the VW01 template at $\lambda_\mathrm{rest}=2770$--$2820$ \AA, the amount of which is $20$\% of the mean flux at $\lambda_\mathrm{rest}=$ 2930--2970 \AA\ \citep{Kurk07}.
Still, the modified VW01 iron template gives a smaller Fe{\sc ii}/Mg{\sc ii} flux ratio of ULAS J1342+0928 than that measured with the T06 template.
This issue is revisited in Section~\ref{sec:discussion_continuum}.

The iron templates were then convolved with Gaussian kernels 
to generate a wide range of velocity dispersion of $\mathrm{FWHM}=500$--$10000$ km s$^{-1}$, with a step of $500$ km s$^{-1}$.
Finally, those templates were redshifted to the systemic redshift of ULAS J1342+0928.
A potential velocity shift of the iron emission lines is not considered in this study.

\subsection{Continuum and Iron Fitting Procedure} \label{sec:spec_model_fit}
The power-law plus Balmer continuum model and the iron templates were iteratively fitted to the observed spectrum.
In this study, the following wavelength ranges were selected as continuum windows:
$\lambda_\mathrm{rest}=$ 1275--1285\AA, 1310--1325\AA, 2500--2750\AA, and 2850--2890\AA.
The first two were chosen to be line-free regions around the O{\sc i}+Si{\sc ii} line at $\lambda_\mathrm{rest}=1305$\AA.
The latter two ranges were also used for fitting the iron templates.
The iron flux at those two wavelength ranges are dominated by Fe{\sc ii} \citep{Vestergaard01}.
The UV Fe{\sc ii} bump actually exists over a wider wavelength region of  $\lambda_\mathrm{rest}\approx$ 2200--3000\AA, but weak non-iron emission lines also exist at $2200<\lambda_\mathrm{rest}<2500$\AA\  such as C{\sc ii]}$\lambda2326$, Fe{\sc iii} UV 47 at $\lambda_\mathrm{rest}\approx2418$\AA\ (not covered by the iron templates), and [O{\sc ii}]$\lambda2470$.

Other wavelength ranges such as $\lambda_\mathrm{rest}=$ 1425--1470\AA,\  1680--1710\AA, and 1975--2050\AA\  were also used in the literature for continuum fitting; however those were excluded for our spectral fitting of ULAS J1342+0928 for the following reasons.
First, ULAS J1342+0928 has extremely broad emission lines with FWHM more than $10,000$ km s$^{-1}$,
and therefore the broad outskirts of the emission lines contribute to the observed flux  between Si{\sc iv} and C{\sc iv}, and also at the redder side of C{\sc iii]}.
Second, weak emission lines are present and heavily blended between C{\sc iv} and C{\sc iii]}: He{\sc ii}$\lambda1640$, O{\sc iii]}$\lambda1663$, N{\sc iv}$\lambda1719$, Al{\sc ii}$\lambda1722$, N{\sc iii]}$\lambda1750$, and Fe{\sc ii} \citep{VB01, Nagao06b}.
In these wavelength regions, the observed flux deviates from the intrinsic power-law continuum due to those weak lines.
Also, the iron templates include Fe{\sc iii} forest at $\lambda_\mathrm{rest}=$ 1900--2100\AA.
This region should be excluded from the continuum and iron windows, as the Fe{\sc ii} line flux is our primary interest.
The continuum flux is overestimated if those wavelength regions are used in the continuum+iron fitting, which ends up underestimating the Fe{\sc ii} flux.

To derive the continuum and iron template parameters, 
an initial continuum model was fitted to the spectrum, after which the iron templates were fitted to the continuum-subtracted residuals.
The continuum model was then re-fitted to the spectrum after subtracting the iron contribution from the original spectrum.
The same continuum and iron templates fitting were repeated until the iron templates fitting achieves a convergence of $<1$\%.
The continuum+iron models were fitted by all the iron templates with different line broadening.
The model that gave the minimum chi-square at the iron windows was adopted as the best model.

Another iterative fitting was done to test potential over-subtraction of the continuum, in which the initial continuum scale was set to $10$\% of the value derived in the continuum-only fitting.
Even in this case, the scales of continuum and the T06 iron template converged at the values that are in agreement with the originally determined ones within $1$\%.
There was no change in the selected widths of Fe{\sc ii}+Fe{\sc iii}.

The best-fit continuum parameters for the T06 model are reported in Table~\ref{tab:continuum}.
Those continuum parameters agree with the VW01 model within $1\sigma$ uncertainty.
The absolute magnitude at rest-frame $1450$\AA\ and monochromatic luminosity at rest-frame $3000$\AA\ ($\lambda L_{3000}$) were measured from the sum of the fluxes of the best-fit power-law and the Balmer continuum models.
The bolometric luminosity $L_\mathrm{bol}$ was derived by applying a bolometric correction of $L_\mathrm{bol}=5.15 \times \lambda L_{3000}$ \citep{Richards06a}.
For the Fe{\sc ii} flux, the scaled iron templates were integrated over $2200 \leq \lambda_\mathrm{rest} \leq 3090$\AA.
The integrated Fe{\sc ii} flux was divided by the continuum flux at $\lambda_\mathrm{rest}=3000$\AA\ to derive the equivalent width. 
This procedure allows for direct comparison with the low-redshift quasars presented in \citet{Sameshima17}, as they derived the Fe{\sc ii} equivalent widths in the same way.

There are metal absorption lines imprinted on the quasar spectrum.
A detailed analysis of those metal absorbers will be presented in elsewhere, and only the $z>6$ Mg{\sc ii} absorbers reported in \citet{Cooper19}\footnote{Those absorbers were identified with their FIRE spectrum, which has a higher spectral resolution than the GNIRS spectrum presented in this paper.} are shown in Figure~\ref{fig:spec_all}.
Those metal absorption lines were automatically masked in the continuum fitting by clipping regions where the observed flux deviates from the best-fit continuum plus emission line model by more than $3\sigma$ of the flux error at each pixel.

\subsection{Emission Lines } \label{sec:spec_model_line}
After subtracting the best-fit continuum (power-law and Balmer continuum) and scaled iron templates, broad emission lines were simultaneously fitted with single Gaussian profiles, from which line widths, equivalent widths, and velocity blueshifts from the [C{\sc ii}] redshift were measured for each line.
The emission lines considered in this paper (except Fe{\sc ii} and Fe{\sc iii}) are Ly$\alpha$+N{\sc v}$\lambda 1240$, Si{\sc ii}$\lambda1263$, Si{\sc iv}$\lambda1397$, C{\sc iv}$\lambda1549$, Al{\sc iii}$\lambda 1857$, C{\sc iii]}$\lambda1909$, and Mg{\sc ii}$\lambda2798$.

In some cases, those emission lines have neighboring weak lines, specifically Si{\sc iv}+O{\sc iv]}$\lambda\lambda1397 \ 1402$ and Si{\sc iii]}+C{\sc iii]}$\lambda\lambda1892\ 1909$.
Since those lines could not be deblended due to their extremely broad nature and the modest spectral resolution,
single Gaussians were fitted to those two multiplets for simplicity.
For C{\sc iii]}, the UV Fe{\sc iii} emission lines at $\lambda_\mathrm{rest}\approx 1900$\AA\ may also contaminate to the C{\sc iii]} flux as Fe{\sc iii} were not directly fitted to the spectrum but just scaled by the same factor of Fe{\sc ii} determined at $\lambda_\mathrm{rest}>2200$\AA.
Al{\sc iii}$\lambda 1857$ and C{\sc iii]}$\lambda 1909$ were simultaneously fitted with two Gaussian profiles.
The two lines were fitted at $\lambda_\mathrm{rest}=$ 1827--1914\AA\ to avoid contamination from adjacent weak lines.
Also, C{\sc iv} was fitted at $\lambda_\mathrm{rest}\leq1570$\AA\ not to be contaminated by the unidentified He{\sc ii}$\lambda1640$.

N{\sc v}$\lambda 1240$ is heavily blended with Ly$\alpha$ and could not be identified as a single line.
For this reason, the line flux of Ly$\alpha$+N{\sc v} composite was measured instead by summing the flux above the power-law continuum at $\lambda_\mathrm{rest}=$ 1160--1290\AA\ \citep{Diamond-Stanic09}.
Note that the flux measured with this procedure includes Si{\sc ii}, which was individually fitted with a single Gaussian profile in this study.
Nevertheless, the same definition of the Ly$\alpha$+N{\sc v} equivalent width was taken for consistency with the literature. 

The uncertainties of the continuum and emission line properties were measured with a Monte Carlo simulation, in which \replaced{100}{1000} mock spectra were generated by adding random noise to the observed spectrum using the flux errors.
Both the continuum and emission line parameters were repeatedly measured for each realization, to take into account the effects of continuum errors on the emission line measurements.
The $1\sigma$ uncertainty range for each quantity was then measured by the 16th and 84th percentiles.
Table~\ref{tab:emission_line} shows the emission line profiles of ULAS J1342+0928 for the T06 model, while the Mg{\sc ii} and Fe{\sc ii} profiles for the VW01 model are also shown in the same table.

\begin{deluxetable*}{CCCCC}[tb!]
\tablecaption{Continuum Properties of ULAS J1342+0928 \label{tab:continuum}}
\tablecolumns{5}
\tablenum{1}
\tablewidth{0pt}
\tablehead{
\colhead{$\alpha_\lambda$} &
\colhead{$F_0$ [$10^{-10}$ erg cm$^{-2}$ s$^{-1}$ \AA$^{-1}$]} &
\colhead{$M_{1450}$ [mag]} &
\colhead{$\lambda L_{3000}$ [$10^{46}$ erg s$^{-1}$]} &
\colhead{$L_\mathrm{bol}$ [$10^{47}$ erg s$^{-1}$]} 
}
\startdata
-1.84_{-0.01}^{+0.03} & 1.60_{-0.37}^{+0.13} & -26.57\pm0.04 & 2.47\pm 0.03 & 1.27\pm0.02 \\
\enddata
\tablecomments{
Those continuum properties are based on the best-fit power-law component of the T06 model. 
The continuum luminosity for the VW01 iron model is identical to the T06 model.
}
\end{deluxetable*}

\begin{deluxetable*}{lCCCCCCCCCC}[htbp!]
\tablecaption{Emission Line Properties of ULAS J1342+0928 \label{tab:emission_line}}
\tablecolumns{11}
\tablenum{2}
\tablewidth{0pt}
\tablehead{
\colhead{} &
\colhead{Ly$\alpha$+N{\sc v}$^*$} &
\colhead{Si{\sc ii}} &
\colhead{Si{\sc iv}} &
\colhead{C{\sc iv}} &
\colhead{Al{\sc iii}} &
\colhead{C{\sc iii]}} &
\colhead{Mg{\sc ii}} &
\colhead{Fe{\sc ii}$^\dagger$} &
\colhead{Mg{\sc ii}$_\mathrm{VW01}$} &
\colhead{Fe{\sc ii}$^\dagger_\mathrm{VW01}$} 
}
\startdata
 FWHM (km s$^{-1}$) &\cdots& 4040_{-290}^{+300} & 10840_{-380}^{+350} & 16000_{-430}^{+390} &5300_{-670}^{+610} & 11880_{-630}^{+470} & 2830_{-210}^{+210} & 1500 & 3780_{-260}^{+220} & 3500 \\
 EW$_\mathrm{rest}$ (\AA) & 12.5_{-0.2}^{+1.3}& 4.3_{-0.3}^{+0.4} & 14.3_{-0.8}^{+0.8} & 23.2_{-1.0}^{+1.0} & 6.7_{-1.0}^{+1.2}& 31.9_{-1.9}^{+1.0} & 13.4_{-0.9}^{+0.8} & 126_{-15}^{+6} & 19.9_{-1.5}^{+0.7} & 123_{-19}^{+4} \\
 $\Delta v_\mathrm{[CII]}$ (km s$^{-1}$) & \cdots & 1340_{-90}^{+100} & 3640_{-130}^{+150} & 5510_{-110}^{+240} & 2630_{-320}^{+330}& 2020_{-190}^{+180} & 340_{-80}^{+110} & \cdots & 750_{-90}^{+90} & \cdots \\
\enddata
\tablecomments{
The line measurements are based on the T06 model, while for Mg{\sc ii} and Fe{\sc ii}, those based on our VW01 model are also reported in the last two columns.
For the other emission lines considered, the two models agree with each other within $2\sigma$.
The line widths are corrected for instrumental broadening.
The positive values in $\Delta v_\mathrm{[C{\sc II}]}$ are blueshifts with respect to [C{\sc ii}].
}
\tablenotetext{*}{The Ly$\alpha$ + N{\sc v} equivalent width includes the Si{\sc ii}$\lambda 1263$ contribution, as the line flux was measured by summing flux above the continuum models at $\lambda_\mathrm{rest}=1160$--$1290$\AA, following the empirical definition of \citet{Diamond-Stanic09}.}
\end{deluxetable*}

\section{Results} \label{sec:spec_result}
Here, the results of the UV spectral modeling of ULAS J1342+0928 are presented based on the T06 model, unless otherwise stated.
\subsection{Continuum Properties} \label{sec:spec_result_cont}
The continuum slope of $\alpha_\lambda=-1.84^{+0.03}_{-0.01}$ is a typical value for lower-redshift quasars \citep[e.g.,][]{VB01, Selsing16}.
In \citet{Banados18}, only the power-law component was considered for their continuum model.
The derived absolute magnitude at rest-frame $1450$\AA\ in our new continuum fitting, $M_{1450}=-26.57\pm0.04$ is fainter than the previously reported value by $\Delta M_{1450}=0.19$ mag.
This difference is mostly attributed to the contribution of iron emission lines, which is taken into account in this work, while the different cosmology used in \citet[][$H_0=67.7 \mathrm{\ km\ s^{-1}\  Mpc^{-1}}$, $\Omega_m=0.307$, $\Omega_\Lambda=0.693$]{Banados18} accounts for a $0.06$ mag of the difference compared to our new measurement.

\subsection{Broad Emission Line Properties} \label{sec:spec_result_line}
\begin{figure}[htb!]
\centering
 \includegraphics[width=\linewidth]{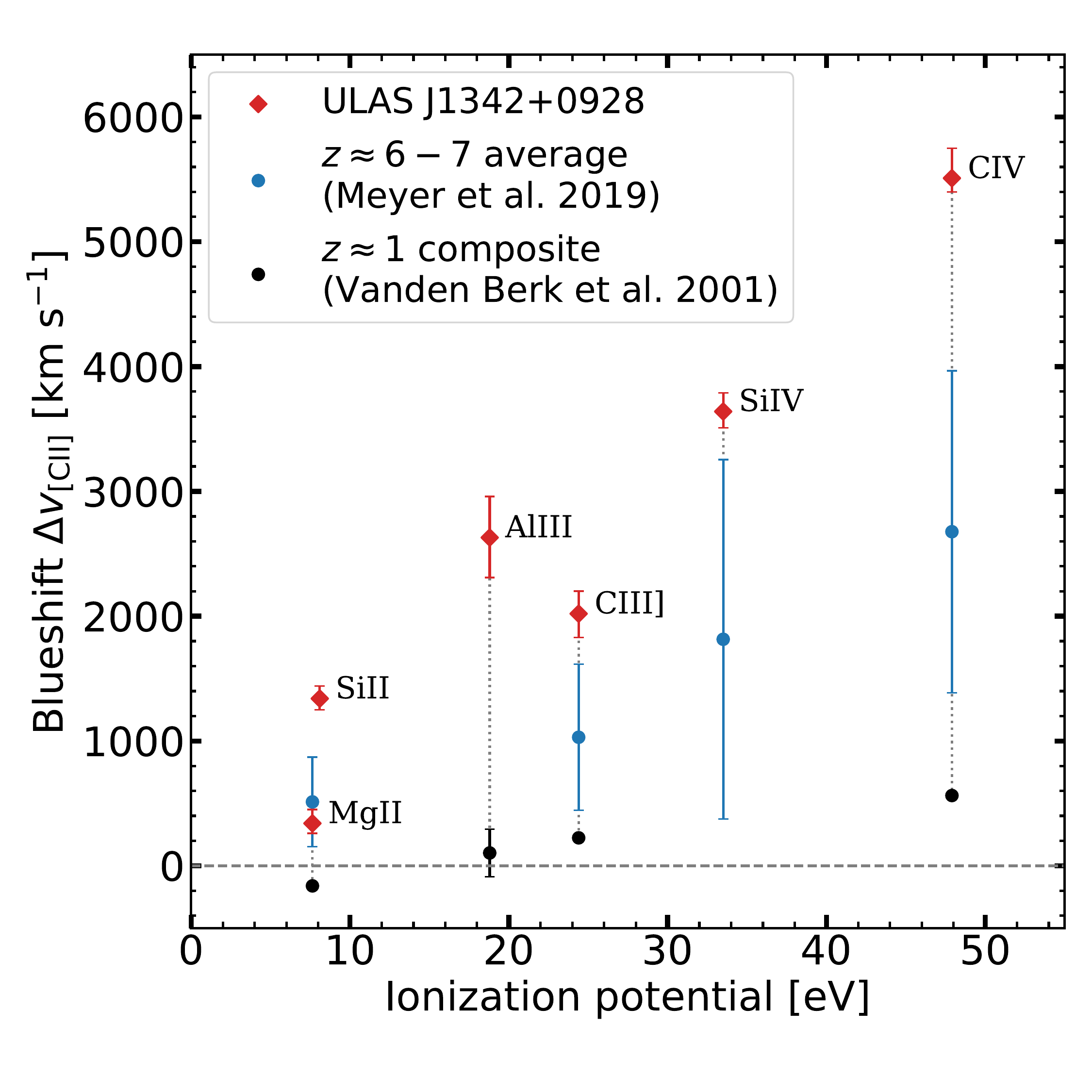}
\caption{
Emission line blueshifts as a function of ionization potentials.
Positive values are blueshifts.
The blueshifts of ULAS J1342+0928 with respect to the [C{\sc ii}] $158\micron$ line are shown in red diamonds with the measurement errors indicated by the errorbars.
The average blueshifts of  $z\approx 6$--$7$ quasars in \citet[][except their measurements for J1342+0928]{Meyer19} are shown in blue circles, with the errorbars showing the standard deviation.
For this measurements, 16 quasars are used for which the rest-FIR [C{\sc ii}] or CO(6-5) redshifts are available in the literature (See text for details).
The blueshifts in the composite spectrum of the low-redshift ($z\approx1$) SDSS quasars with respect to [O{\sc iii}] $\lambda5007$ \citep{VB01} are shown in black circles.
Note that Si{\sc ii} and Si{\sc iv} were identified in \citet{VB01} but were excluded in their blueshift measurements, because they took into consideration the line blending of Si{\sc ii} with Ly$\alpha$+N{\sc v} and of Si{\sc iv} with O{\sc iv]}.
} \label{fig:blueshift}
\end{figure}

ULAS J1342+0928 is characterized by highly blueshifted and broad high-ionization lines.
Figure~\ref{fig:blueshift} shows the line blueshifts for the broad emission lines detected in ULAS J1342+0928 as a function of ionization potentials.
The amounts of blueshifts are proportional to the ionization potentials, with the high-ionization lines such as  C{\sc iv} (ionization potential: $47.9$eV) and Si{\sc iv} ($33.5$eV) showing more than $3000$ km s$^{-1}$ blueshifts with respect to the [C{\sc ii}] redshift ($\Delta v_\mathrm{[CII]}=5510_{-110}^{+240}$ km s$^{-1}$ and $3640_{-130}^{+150}$ km s$^{-1}$, respectively).
The correlation between the ionization potentials and the blueshifts was reported in \citet{VB01}, in which they showed a composite spectrum of $z\approx1$ quasars from the Sloan Digital Sky Survey (SDSS, \citealt{York00}).
The blueshifts of the SDSS $z\approx1$ quasars with respect to the forbidden [O{\sc iii}] $\lambda 5007$ line are much smaller than those found in ULAS J1342+0928  for the same lines ($<560$ km s$^{-1}$, see Fig.~\ref{fig:blueshift})\footnote{The narrow [O{\sc iii}]$\lambda5007$ line is known to trace the systemic redshift well (with the typical blueshift of $40$ km s$^{-1}$), while in some cases (e.g., high Eddington ratios) this could show significant blueshifts up to $400$ km s$^{-1}$  \citep[e.g.,][]{Boroson05}.}.

\begin{figure}[htb!]
\centering
 \includegraphics[width=\linewidth]{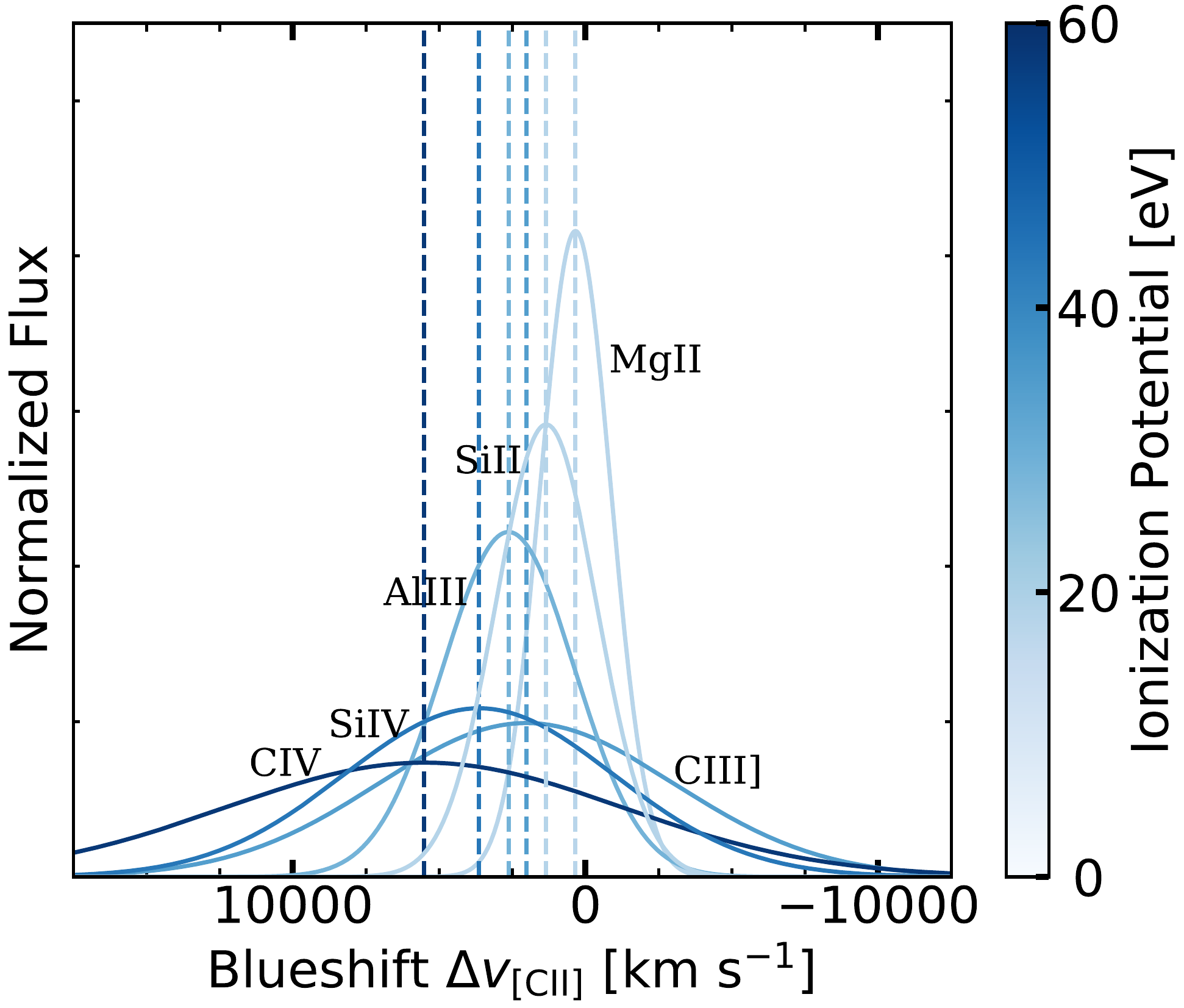}
\caption{
The comparison of the best-fit Gaussian profiles of the detected broad emission lines in the velocity space.
The velocity zeropoint is set to the systemic [C{\sc ii}] redshift from \citet{Banados19}.
The line strengths are scaled so that the integrated fluxes are the same.
The line ionization potentials are indicated with the line colors (higher-ionization lines have darker colors).
The line centers are indicated with dashed lines.
As is also partly seen in Figure~\ref{fig:blueshift}, there is a tendency that higher ionization lines have larger blueshifts and wider profiles.
} \label{fig:blueshift2_2}
\end{figure}

The blueshifts of ULAS J1342+0928 are even larger than those of quasars at similar redshifts.
The composite spectrum of $z\sim6$ luminous quasars from \citet{Shen19} is over-plotted on the ULAS J1342+0928's spectrum in Figure~\ref{fig:spec_all}, where it is evident that ULAS J1342+0928 has larger blueshifts than the average of $z\sim6$ quasars.
\citet{Meyer19} argue that, with their compilation of luminosity-matched $1.5\lesssim z\lesssim7.5$ quasars including ULAS J1342+0928, 
the C{\sc iv} blueshift sharply increases at $z\gtrsim6$.
In \citet{Meyer19}, there are 16 quasars the far-infrared (FIR) [C{\sc ii}] and/or CO(6-5) lines of which were detected in the literature \citep{Carilli07,Wang13,Banados15b,Wang16,Venemans16,Venemans17a,Mazzucchelli17,Decarli18,Venemans19}.
The average blueshifts with respect to the systemic redshifts from the rest-FIR emission lines\footnote{[C{\sc ii}] redshift was prioritized when both the [C{\sc ii}] and CO(6-5) redshifts are available.}, as well as their standard deviation, are shown in Figure~\ref{fig:blueshift}.
For  Mg{\sc ii}, C{\sc iii]}, Si{\sc iv}, and C{\sc iv}, ULAS J1342+0928 has similar or higher blueshifts than this average by a factor of up to $2$, indicating extreme outflowing components in the BLR clouds.

High-ionization emission lines are easily affected by the BLR outflow, showing $\Delta v \approx$ 1000--2000 km s$^{-1}$ velocity shifts in general \citep[e.g.,][]{Richards11, Shen16, Ge19}.
Figure~\ref{fig:blueshift2_2} compares the best-fit Gaussian profiles of those emission lines in the velocity space with the lines color-coded by the ionization potentials.
There is a trend of wider widths in more largely blue-shifted (or higher ionization) emission lines up to FWHM$=16000$ km s$^{-1}$, which is seen in C{\sc iv}.
This is perhaps a natural consequence of the highly blueshifted emission lines that likely have both virialized and outflowing components \citep{Vietri18}, albeit the broad emission lines of ULAS J1342+0928 are well modeled with single Gaussian profiles.
\citet{Vietri18} found that the most luminous class of $2<z<4$ quasars at $L_\mathrm{bol}>2\times10^{47}$ erg s$^{-1}$, which is slightly brighter than ULAS J1342+0928, show similarly large blueshifts up to $8000$ km s$^{-1}$.
ULAS J1342+0928 falls at a similar location as their samples in the C{\sc iv} blueshift - equivalent width plane \citep{Richards11, Plotkin15, Coatman17}.

It is also remarkable that 
Mg{\sc ii}, the lowest-ionization line among the detected emission lines with the ionization potential of $7.6$eV, has a modest blueshift of $\Delta v_\mathrm{MgII-[CII]}=340_{-80}^{+110}$ km s$^{-1}$.
A larger value was reported in \citet[][$\Delta v_\mathrm{MgII-[CII]}=500\pm140$ km s$^{-1}$]{Banados18}, where only the power-law continuum was considered in the spectral modeling (See Section~\ref{sec:spec_model} for the details of our modeling in this paper).
The same trend has also been reported in other luminous quasars at $z\gtrsim6$ \citep{Wang16, Venemans16, Mazzucchelli17}.
\citet{Venemans16} showed similar Mg{\sc ii} blueshifts with respect to [C{\sc ii}] and/or CO(6-5) for their 11 quasars at $z\gtrsim6$ (their mean and standard deviation are $\Delta v_\mathrm{MgII-[CII]/CO}=480\pm630$ km s$^{-1}$).

 The Ly$\alpha$+N{\sc v} equivalent width $\mathrm{EW(Ly\alpha+N{\sc v})_{rest}}=12.5^{+1.3}_{-0.2}\mathrm{\AA}$ classifies ULAS J1342+0928 as a weak-line quasar according to an empirical definition of \citet[][EW(Ly$\alpha$+N{\sc v})$_\mathrm{rest}<15.4$\AA]{Diamond-Stanic09}, which is based on 3000 quasars at $3<z<5$.
 This value is also smaller than the mean property of $z\sim6$ quasars compiled in \citet{Banados16}, which shows  $\left<\mathrm{EW(Ly\alpha+N{\sc v})_{rest}}\right> \geq 34.7$\AA\ ($1\sigma$ range of $14.2$--$85.7$\AA) for their compilation of $117$ quasars at $z>5.6$.
 However, the weak-line nature of Ly$\alpha$+N{\sc v} is likely due to strong absorption by the intergalactic medium (IGM) in the line of sight.
 The IGM neutral fraction at $z>7$ is so strong that the Ly$\alpha$ damping wing suppresses the quasar emission even redward of Ly$\alpha$ \citep{Banados18, Davies18}.
 To support this argument, the C{\sc iv} equivalent width of ULAS J1342+0928, EW(C{\sc iv})$_\mathrm{rest}=23.2^{+1.0}_{-1.0}$\AA, is just close to the $1\sigma$ range reported in \citet[][$1\sigma$ range: $26.1$--$67.4 \mathrm{\AA}$]{Diamond-Stanic09}, even though \citet{Shen19} showed that the weak-line quasar fraction with EW(C{\sc iv})$_\mathrm{rest}<10\mathrm{\AA}$ increases towards $z\sim6$.
 Therefore, it is likely that the weak Ly$\alpha$ is not an intrinsic characteristic of ULAS J1342+0928, and the inherent strength of the BLR emission lines are similar to those of lower-redshift quasars.

\subsection{Line Ratios} \label{sec:spec_result_line_ratio}
The flux ratios of the BLR emission lines are used to test if ULAS J1342+0928 has a metal-poor BLR when the universe was 680 million years old.
Table~\ref{tab:emission_line_ratio} shows the line ratios of Si{\sc ii}/C{\sc iv}, Si{\sc iv}/C{\sc iv}, Al{\sc iii}/C{\sc iv}, C{\sc iii]}/C{\sc iv}, and Fe{\sc ii}/Mg{\sc ii}.
Those line combinations (except Fe{\sc ii}/Mg{\sc ii}) are in fact second-order tracers of the BLR metallicity, while other line combinations such as N{\sc v}/He{\sc ii} and N{\sc v}/C{\sc iv} were known as the better diagnostics \citep[e.g.,][]{Hamann93, Hamann02, Dietrich03a, Nagao06b}.
However, neither N{\sc v} nor He{\sc ii} were identified in ULAS J1342+0928 due to their weakness, heavy line blending and/or atmospheric absorption (Figure~\ref{fig:spec_all}).
Also, the red tail of the damping wing may suppress N{\sc v} \citep{Banados18, Davies18}.

\begin{figure*}[htb!]
\centering
 \includegraphics[width=\linewidth]{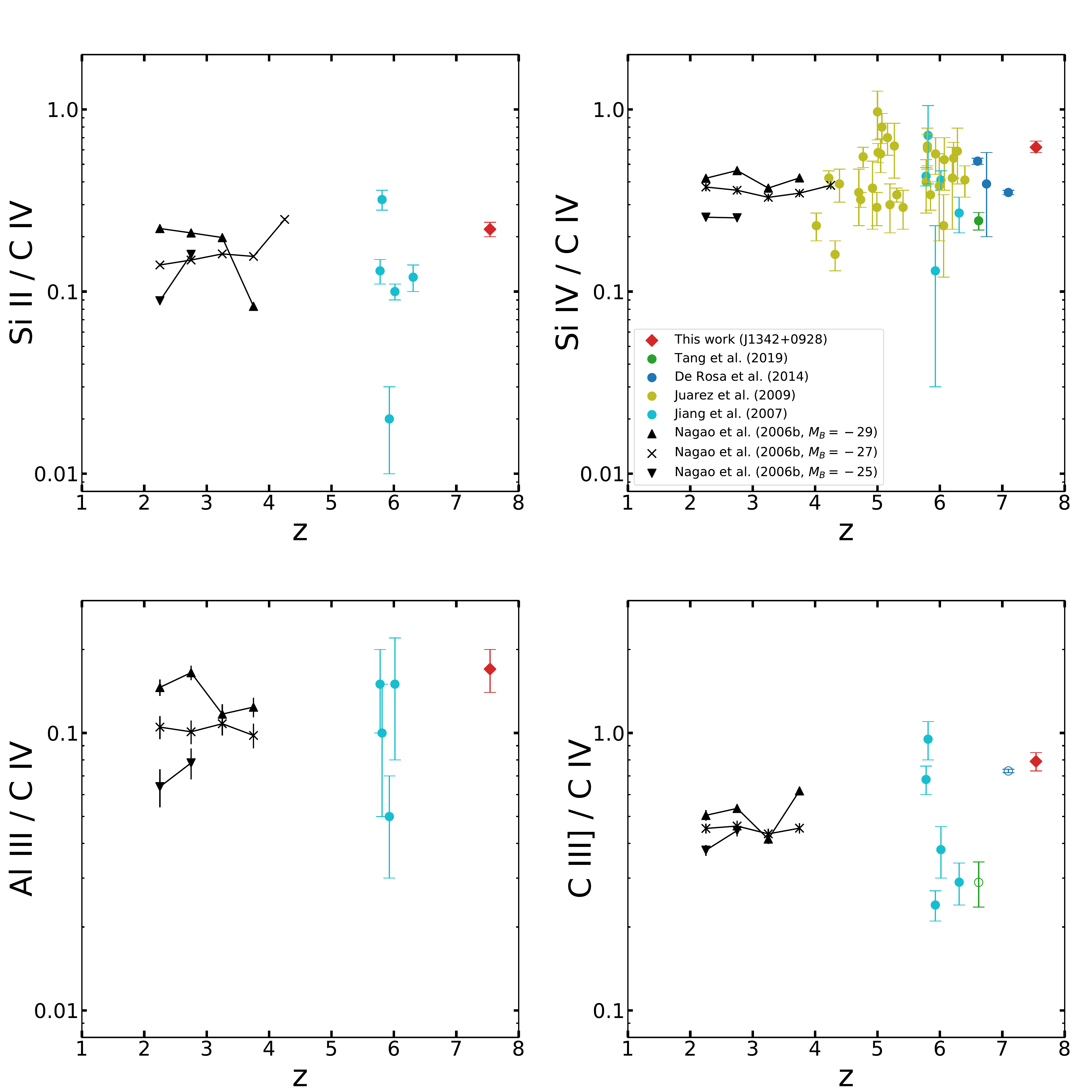}
\caption{
Emission line ratios as a function of redshift.
The colors and symbols are the same in the four panels:
Red: J1342+0928 (T06 model),
blue: \citet{DeRosa14},
green: \citet{Tang19},
cyan: \citet{Jiang07},
yellow: \citet{Juarez09}, and 
black: \citet{Nagao06b}.
For the low-redshift measurements of \citet{Nagao06b}, 
their measurements for three different magnitude ranges ($-29.5\leq M_B\leq -28.5$, $-27.5\leq M_B\leq -26.5$, $-25.5\leq M_B\leq -24.5$) are separately plotted to show that brighter quasars in general show larger flux ratios.
The Si{\sc iii]} $\lambda1892$ flux is added to the datapoints of \citet{Nagao06b} for the C{\sc iii]}/C{\sc iv} flux ratio (bottom right panel), because this weak line was not deblended in the other studies that are quoted in this figure.
\citet{DeRosa14} and \citet{Tang19} are shown as open circles in the C{\sc iii]}/C{\sc iv} panel, because those two studies fitted Al{\sc iii}+Si{\sc iii]}+C{\sc iii]} as a single line. 
} \label{fig:SiIV_CIV}
\end{figure*}

Figure~\ref{fig:SiIV_CIV} shows the line ratios as a function of redshift.
The lower-redshift datapoints come from the literature: 
\citet{Jiang07}, \citet{DeRosa14}, and \citet{Tang19} for $6\lesssim z\lesssim7$, \citet{Juarez09} for $4\lesssim z\lesssim6$, and \citet{Nagao06b} for $2\lesssim z\lesssim4$.
Since there is a magnitude dependence in the BLR line ratios,
the subsets of the $2\lesssim z\lesssim4$ measurements are shown with different $B$-band magnitude bins, namely $-29.5\leq M_B \leq -28.5$,  $-27.5\leq M_B \leq -26.5$, and  $-25.5\leq M_B\leq-24.5$.

\begin{deluxetable*}{lCCCCC}[htbp!]
\tablecaption{Emission Line Ratio \label{tab:emission_line_ratio}}
\tablecolumns{6}
\tablenum{3}
\tablewidth{0pt}
\tablehead{
\colhead{Continuum model} &
\colhead{Si{\sc ii}/C{\sc iv}} &
\colhead{Si{\sc iv}/C{\sc iv}} &
\colhead{Al{\sc iii}/C{\sc iv}} &
\colhead{C{\sc iii]}/C{\sc iv}} &
\colhead{Fe{\sc ii}/Mg{\sc ii}} 
}
\startdata
T06 & 0.22_{-0.02}^{+0.02} & 0.62_{-0.05}^{+0.04} & 0.17_{-0.03}^{+0.03} & 0.79_{-0.06}^{+0.04} & 8.7_{-1.3}^{+0.6}  \\ \hline
VW01 & 0.22_{-0.02}^{+0.02} & 0.62_{-0.04}^{+0.04} &0.17_{-0.03}^{+0.02} & 0.78_{-0.05
}^{+0.04} & 5.5_{-0.8}^{+0.4}  \\ 
T06\_noBC & 0.26_{-0.02}^{+0.02} & 0.66_{-0.05}^{+0.05} & 0.17_{-0.03}^{+0.04} & 0.75_{-0.06}^{+0.05} & 7.7_{-0.9}^{+0.6}  \\ 
T06\_wide & 0.19_{-0.01}^{+0.02} & 0.55_{-0.03}^{+0.03} & 0.16_{-0.02}^{+0.03} & 0.71_{-0.04}^{+0.04} & 1.4_{-0.6}^{+0.4}  \\
\enddata
\tablecomments{
The emission line ratios are shown for four different models: ``T06" and ``VW01" denote the (power-law and Balmer) continuum+iron+emission line fits in which the iron templates of \citet{Tsuzuki06} and \citet{Vestergaard01} are used, respectively.
``T06\_noBC"  denotes the model in which the \citet{Tsuzuki06} template was used but only the power-law continuum was fitted for the continuum component.
``T06\_wide" is the model in which the spectrum is fitted with the same components as those of the T06 model but with a wider continuum+iron window (see Section~\ref{sec:discussion_continuum} for more details).
}
\end{deluxetable*}

Those emission line ratios have a positive correlation with the BLR metallicity.
Therefore, our result suggests that J1342+0928 has a comparable BLR metallicity to the lower-redshift quasars.
This result extends the no redshift evolution trend of the BLR metallicity up to $z=7.5$.
The magnitude dependence of the emission line ratios does not affect our conclusion, because the $4\lesssim z\lesssim 7$ quasars shown in Figure~\ref{fig:SiIV_CIV} are primarily at the most luminous range ($L_\mathrm{bol}\gtrsim10^{46}$ erg s$^{-1}$).

It is important to recall that Si{\sc iv} and C{\sc iii]} have adjacent weak lines (i.e., O{\sc iv]} $\lambda1402$ and Si{\sc iii]}$\lambda1892$) that were ignored in the line fitting and were therefore not deblended.
Specifically, O{\sc iv]} may contribute to the observed Si{\sc iv}/C{\sc iv} line ratio, although the previous studies plotted in the top right panel of Figure~\ref{fig:SiIV_CIV} also modeled the blended Si{\sc iv}+O{\sc iv]} line as a single line.
For Al{\sc iii}+Si{\sc iii]}+C{\sc iii]}, \citet{Nagao06b} were able to deblend the three lines, while \citet{DeRosa14} and \citet{Tang19} assumed the multiplet line as a single line.
\citet{Jiang16} deblended Al{\sc iii} and C{\sc iii]}, and ignored Si{\sc iii]} as did this study.
In Figure~\ref{fig:SiIV_CIV}, Al{\sc iii}/C{\sc iv} line ratios are shown in the case that Al{\sc iii} was measured as a single line.
For C{\sc iii]}, the sum of C{\sc iii]} and Si{\sc iii]} are shown for the measurements of \citet{Nagao06b}, to be consistent with \citet{Jiang07} and this study.
Although Al{\sc iii} was not deblended, the C{\sc iii]}/C{\sc iv} measurements of \citet{DeRosa14} and \citet{Tang19} are also shown in Figure~\ref{fig:SiIV_CIV} for reference.
In fact, the relative strength of Si{\sc iii]} with respect to C{\sc iii]} is controversial, and this third line could add significant flux around C{\sc iii]} (See \citealt[][Sec~4.1]{Nagao06b}). 
Moreover, there could be a non-negligible contribution of the Fe{\sc iii} multiplet at $\lambda_\mathrm{rest}\approx1900$\AA\ \citep{VB01, Vestergaard01}, which was just considered by scaling the iron templates with the factor determined at the Fe{\sc ii} multiplet at $\lambda_\mathrm{rest}=$ 2500--2890\AA.
Therefore, it is not clear if the apparently high C{\sc iii]}/C{\sc iv} line ratio and the large scatter at $z\gtrsim6$  (as was also pointed out by \citealt{DeRosa14}) reflect  intrinsic changes in the C{\sc iii]} emission line, or due to the complex line blending of C{\sc iii]} with Al{\sc iii}, Si{\sc iii]} and Fe{\sc iii}.

\citet{Venemans17} and \citet{Novak19} investigated
the host galaxy properties of ULAS J1342+0928 by rest-frame FIR dust continuum and atomic emission lines from the interstellar medium (ISM).
They revealed that the host galaxy has a high star-formation rate ($\sim150 M_\odot \mathrm{yr^{-1}}$) and that its ISM is rich in dust ($M_\mathrm{dust}=3.5\times10^7 M_\odot$).
The line ratio of the rest-frame FIR [O{\sc iii}] $88\micron$ and [N{\sc ii}] $122\micron$ emission lines suggested that the ISM gas-phase metallicity is consistent with the solar value.
Therefore, ULAS J1342+0928 experienced rapid metal enrichment both over the BLR (sub-pc) and the ISM (kpc) scales.

\subsection{Fe{\sc ii}/Mg{\sc ii}} \label{sec:iron_abundance}
\begin{figure*}[htb!]
\centering
 \includegraphics[width=\linewidth]{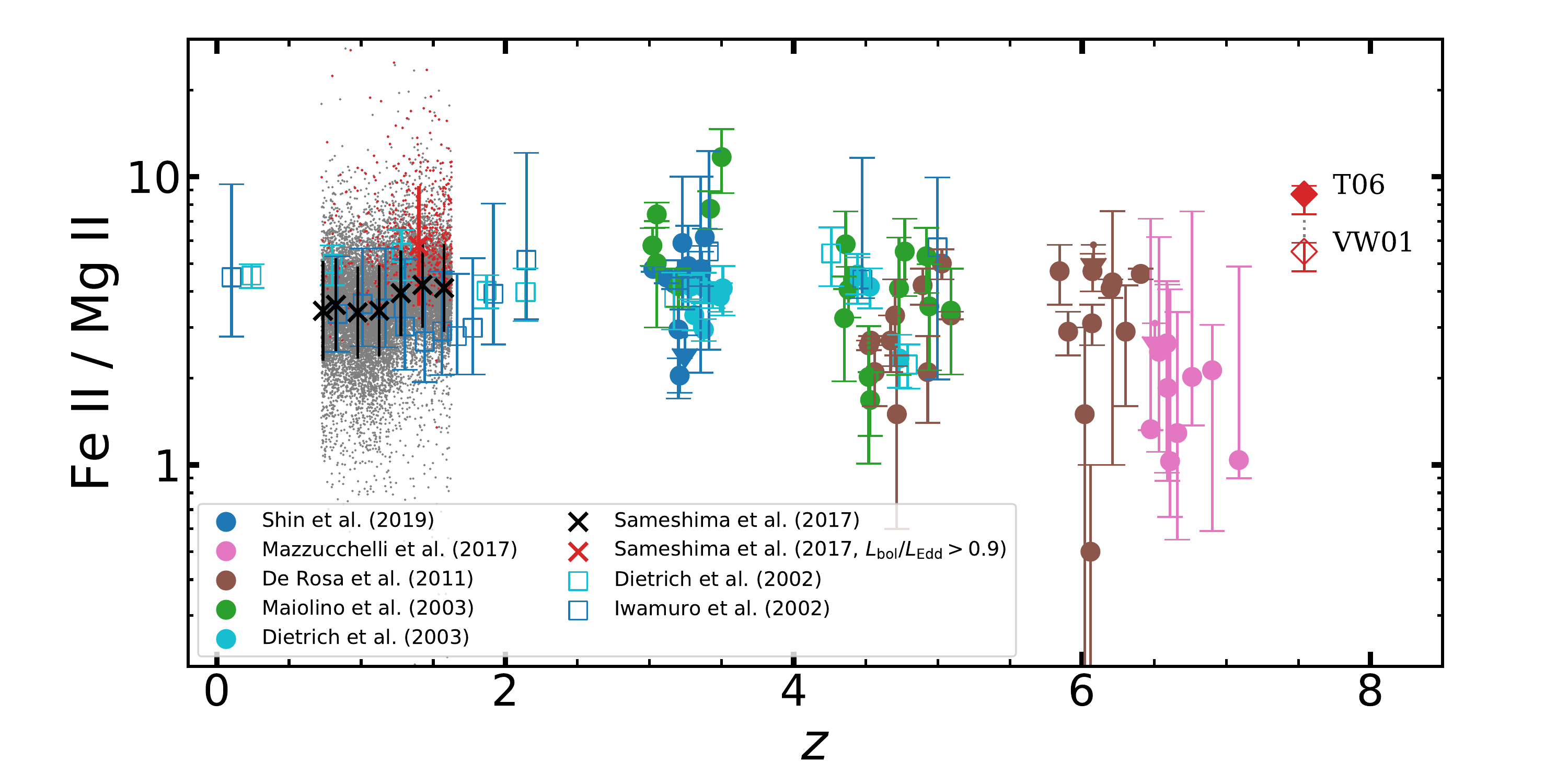}
\caption{
Fe{\sc ii}/Mg{\sc ii} line ratios as a function of redshift.
Red diamonds show ULAS J1342+0928 with the different symbols corresponding to different iron templates used in the spectral decomposition (filled symbol: T06, open symbol: VW01).
The colored circles show the  Fe{\sc ii}/Mg{\sc ii} measurements of individual quasars in the literature: 
\citet[][cyan]{Dietrich03}, \citet[][green]{Maiolino03}, \citet[][brown]{DeRosa11}, 
\citet[][magenta]{Mazzucchelli17}, and \citet[][blue]{Shin19}.
The SDSS DR7 quasars at $0.72<z<1.63$  from \citet{Sameshima17} are shown as dots.
Their median Fe{\sc ii}/Mg{\sc ii} flux ratios binned by redshift ($\Delta z=0.15$) are shown as black crosses with the error bars showing standard deviation.
The low-redshift SDSS quasars with the Eddington ratios over $L_\mathrm{bol}/L_\mathrm{Edd}=0.9$ (comparable to ULAS J1342+0928) are shown in red dots, with the median of this sub sample shown in a red cross.
Open squares show the redshift evolution down to $z\sim5$ based on the median values compiled by \citet[][blue]{Iwamuro02} and composite spectra constructed by \citet[][cyan]{Dietrich02b}.
For the latter measurement, the  Fe{\sc ii}/Mg{\sc ii} values presented in \citet{Dietrich03} are quoted.
} \label{fig:iron_zevolution}
\end{figure*}

The Fe{\sc ii}/Mg{\sc ii} flux ratio has been used as a proxy for the Fe/$\alpha$-element abundance ratios in the BLR clouds.
The Fe{\sc ii}/Mg{\sc ii} line ratio of ULAS J342+0928 is Fe{\sc ii}/Mg{\sc ii} $=8.7_{-1.3}^{+0.6}$ (for the T06 model)
as reported in Table~\ref{tab:emission_line_ratio}.
The VW01 model gives a smaller value with Fe{\sc ii}/Mg{\sc ii} $=5.5_{-0.8}^{+0.4}$.
This difference is due to the overestimated Mg{\sc ii} flux in the VW01 iron template, as  revisited in Section~\ref{sec:discussion_continuum}.
In Figure~\ref{fig:iron_zevolution}, the Fe{\sc ii}/Mg{\sc ii} measurements in the literature are compiled as a function of redshift,
in which both the T06 and VW01 models are shown for ULAS J1342+0928.
The filled circles at $3\lesssim z\lesssim7$ show the measurements of individual quasars \citep{Maiolino03,Dietrich03,DeRosa11,Mazzucchelli17,Shin19}.
Since the same quasars are frequently used in more than one paper,
only the most recent measurements are shown in Figure~\ref{fig:iron_zevolution} to avoid sample duplication. 
For the low-redshift quasars, the Fe{\sc ii}/Mg{\sc ii} values of the individual SDSS DR7 quasars from \citet{Sameshima17} are shown with their median values binned by  redshift (with a step of $\Delta z=0.15$).
The redshift average over $0.1<z<5.0$ reported in \citet{Iwamuro02} and the  Fe{\sc ii}/Mg{\sc ii} values of the composite spectra at $0.2\leq z\leq4.8$ that were constructed by \citet{Dietrich02b} are compared with the SDSS quasars in Figure~\ref{fig:iron_zevolution}.

ULAS J1342+0928 is apparently as iron-rich as the lower-redshift quasars, as its Fe{\sc ii}/Mg{\sc ii} line ratio is comparable to the $z\lesssim7$ quasars.
However, comparison with those measurements in the literature is not simple, because they often use different methods to model the quasar continuum and iron forests.
For example, many of the  $z\gtrsim6$ quasars were measured with the VW01 iron template \citep{Mazzucchelli17, DeRosa11, Jiang07}, while 
\citet{Sameshima17} use the T06 template for their low-redshift quasars.
\citet{Shin19} applied both the VW01 and T06 templates\footnote{
Their latter model is quoted in Figure~\ref{fig:iron_zevolution} to be consistent with our measurement of ULAS J1342+0928.}.
The Fe{\sc ii}/Mg{\sc ii} line ratio gets $0.20$ dex smaller when the \citet{Vestergaard01} model is applied to ULAS J1342+0928 (Fe{\sc ii}/Mg{\sc ii} $=5.5_{-0.8}^{+0.4}$), in which case the line ratio is closer to the $z\gtrsim6.5$ quasars in \citet{Mazzucchelli17}.
The systematic uncertainty in the Fe{\sc ii}/Mg{\sc ii} measurement is further discussed in Section~\ref{sec:discussion_continuum}.
In either case, our deep observation of ULAS J1342+0928 suggests rapid iron enrichment in the BLR clouds at $z>7.54$.

Moreover, photoionization calculations suggest that the strengths of the Fe{\sc ii} and the Mg{\sc ii} emission lines are affected by non-abundance parameters, such as the BLR gas density and the microturbulence within the gas clouds \citep[e.g.,][]{Verner99,Baldwin04, Dong11, Sameshima17}.
\citet{Sameshima17} showed that the gas density is the dominant non-abundance parameter that changes the Fe{\sc ii}/Mg{\sc ii} line ratio.
This argument is based on the fact in their photoionization model that the two emission lines originate from different regions inside the BLR gas clouds.

\begin{figure*}[htb!]
\centering
 \includegraphics[width=\linewidth]{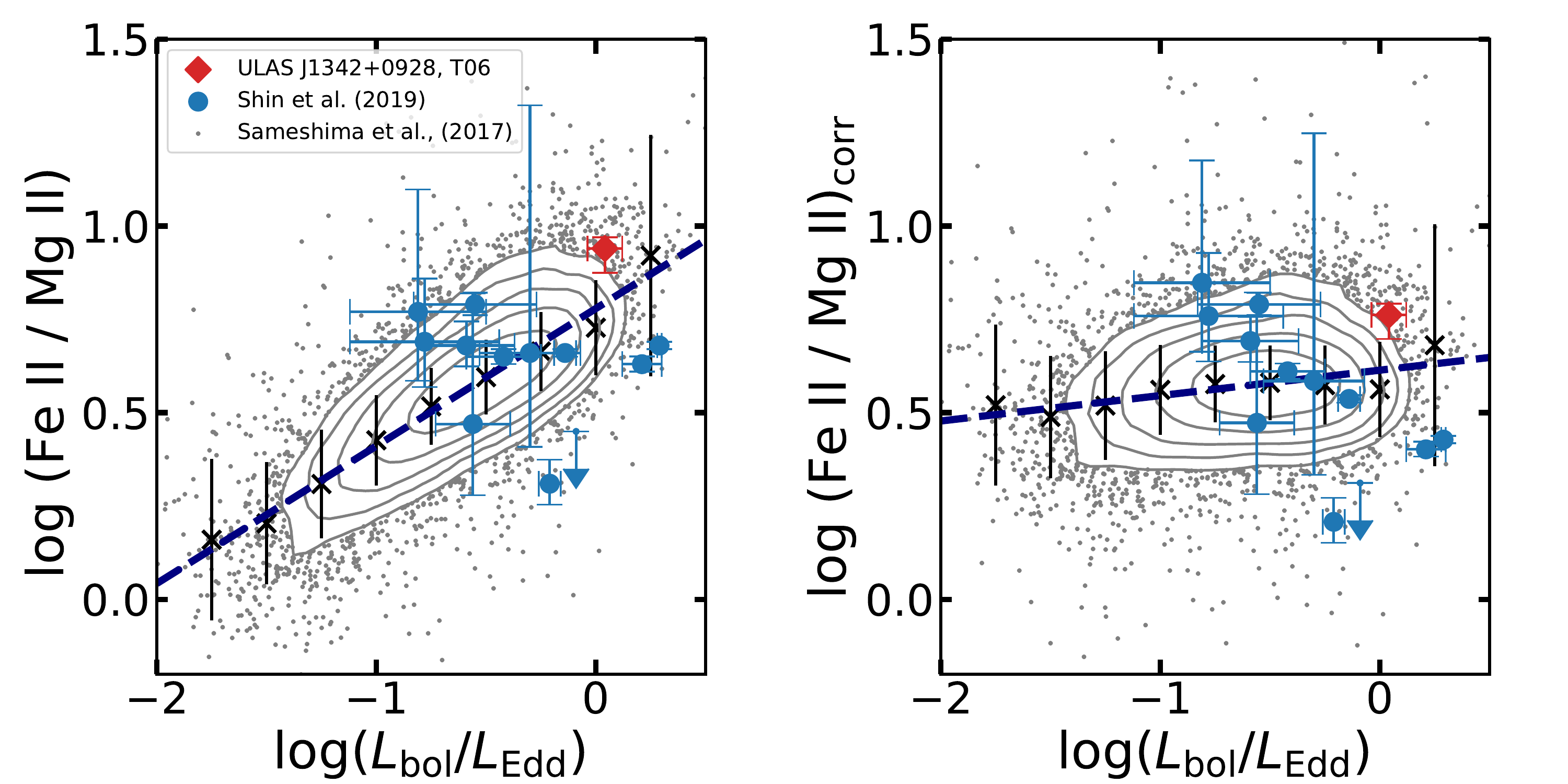}
\caption{
{\it Left:} The dependence of the Fe{\sc ii}/Mg{\sc ii} line ratio on the Eddington ratio.
Symbols are the same as Figure~\ref{fig:iron_zevolution},
while only the measurements using \citet{Tsuzuki06} iron template are shown in this plot.
The lognormal local density of the SDSS DR7 quasars are shown in grey contours with a grid size of $0.3$ dex.
For the SDSS DR7 quasars, their median and standard deviation at small bins ($\Delta \log{\left(L_\mathrm{bol}/L_\mathrm{Edd}\right)}=0.25$ dex) are also shown in crosses, with the dashed line being the linear regression to the distribution \citep{Sameshima17}. 
{\it Right:} The same plot with the Fe{\sc ii}/Mg{\sc ii} line ratio corrected for the Eddington ratio dependence given in \citet{Sameshima17}: $\log \left(\mathrm{Fe{\sc II}}/\mathrm{Mg{\sc II}}\right)_\mathrm{corr}=\log \left(\mathrm{Fe{\sc II}}/\mathrm{Mg{\sc II}}\right)-0.30 (\log \left(L_\mathrm{bol}/L_\mathrm{Edd}\right)+0.55)$.
} \label{fig:iron_Edd}
\end{figure*}

The left panel of Figure~\ref{fig:iron_Edd} shows the  Fe{\sc ii}/Mg{\sc ii} - $L_\mathrm{bol}/L_\mathrm{Edd}$ plane, in which only the measurements using the T06 iron template are shown (i.e., \citealt{Sameshima17},  \citealt{Shin19}, and this study) to avoid scatters due to different choice of iron templates.
In this figure, it is clear that the low-redshift SDSS quasars have a positive correlation between Fe{\sc ii}/Mg{\sc ii} and the Eddington ratio.
While ULAS J1342+0928 has an approximately twice higher Fe{\sc ii}/Mg{\sc ii} ratio than the median value of the entire low-redshift quasars, 
the $z<2$ SDSS quasars at the near or super-Eddington range ($L_\mathrm{bol}/L_\mathrm{Edd}>0.9$) have similar Fe{\sc ii}/Mg{\sc ii} ratios to ULAS J1342+0928.
Those low-redshift counterparts are also shown in red dots in Figure~\ref{fig:iron_zevolution}.
In order to better trace the relative iron abundance over magnesium (i.e., $\alpha$-element) in the BLR clouds, \citet{Sameshima17} introduced the correction to the Eddington ratio dependence of the Fe{\sc ii}/Mg{\sc ii} ratios:
\begin{equation}
     \mathrm{EW(Mg{\sc II})}_\mathrm{corr}=\mathrm{EW(Mg{\sc II})}\left(\frac{\left<L_\mathrm{bol}/L_\mathrm{Edd}\right>}{L_\mathrm{bol}/L_\mathrm{Edd}}\right)^{-0.30},
 \end{equation}
 where the normalization factor is $\left<L_\mathrm{bol}/L_\mathrm{Edd}\right>=10^{-0.55}$.
 The corrected Fe{\sc ii}/Mg{\sc ii} - $L_\mathrm{bol}/L_\mathrm{Edd}$ plane in the right panel of Figure~\ref{fig:iron_Edd} has a smaller scatter over the entire Eddington ratio range.
 ULAS J1342+0928 with the corrected (Fe{\sc ii}/Mg{\sc ii})$_\mathrm{corr}=5.8_{-0.9}^{+0.4}$ now has a less extreme Fe{\sc ii}/Mg{\sc ii} with only a $1.6\sigma$ excess from the median of the entire low-redshift SDSS quasars.
Therefore, this correction gives a strong evidence that the iron enrichment is completed at $z>7.54$ (i.e., when the universe is less than $680$ million years old) at least in the scale where the local chemical abundance is represented by the BLR clouds.
Further discussions are provided in Section~\ref{sec:discussion_Fe} about the early iron enrichment.

\subsection{BH mass} \label{secion:MBH}
Our new Mg{\sc ii} line measurement enabled us to revisit the black hole mass of ULAS J1342+0928.
The scaling relation given in \citet{Vestergaard09} was used as in our previous measurement \citep{Banados18}.
The virial mass was derived with the following equation:
\begin{equation}
M_\mathrm{BH} = 10^{6.86} \left(\frac{\mathrm{FWHM(Mg{\sc II})}}{10^3\ \mathrm{km\ s^{-1}}}\right)^2 \left(\frac{\lambda L_\lambda\ (3000\mathrm{\AA})}{10^{44}\ \mathrm{erg\ s^{-1}}}\right)^{0.5}M_\odot,\label{eq:VW01}
\end{equation}
where FWHM(Mg{\sc ii}) is the full width at half maximum of the Mg{\sc ii} line and $\lambda L_\lambda\ (3000\mathrm{\AA})$ is the monochromatic luminosity at rest-frame $3000$\AA.
The measurement uncertainty of the virial black hole mass was derived by propagating the measurement errors of the Mg{\sc ii} line width and the monochromatic luminosity.
The systematic errors are usually larger than the measurement errors of the black hole masses, as \citet{Shen13_review} argue that there is a $0.5$ dex uncertainty in the Mg{\sc ii}-based mass measurements.
A subsequently derived quantity after the $M_\mathrm{BH}$ measurement is the Eddington luminosity:
\begin{equation}
L_\mathrm{Edd}=1.3\times10^{38} \left(\frac{M_\mathrm{BH}}{M_\odot} \right)\ \mathrm{erg\ s^{-1}}.
\end{equation}
The Eddington ratio $L_\mathrm{bol}/L_\mathrm{Edd}$ was derived by dividing the bolometric luminosity by the Eddington luminosity.

The resulting black hole mass for ULAS J1342+0928 is $M_\mathrm{BH}=9.1^{+1.3}_{-1.4} \times10^8 M_\odot$ for the T06 model.
The Eddington ratio is $L_\mathrm{bol}/L_\mathrm{Edd}=1.1^{+0.2}_{-0.2}$.
Only the measurement errors are taken into account in the reported uncertainties.
The Mg{\sc ii}-based black hole mass slightly increases from the previous measurement of $M_\mathrm{BH}=7.6^{+3.2}_{-1.9} \times10^8 M_\odot$ \citep{Banados18}, while the difference is within the $1\sigma$ level of the previous measurement.
Accordingly, the Eddington ratio decreases from $L_\mathrm{bol}/L_\mathrm{Edd}=1.5^{+0.5}_{-0.4}$ to a near Eddington limit.
The reason for this difference, albeit small, is due to the wider Mg{\sc ii} width and the suppressed continuum luminosity by the additional iron pseudo-continuum considered in the spectral fitting.

The virial black hole mass becomes even more massive when the VW01 model is used.
As reported in Table~\ref{tab:emission_line}, the Mg{\sc ii} width $34$\% broader than that in the T06 model.
The broader line profile increases the Mg{\sc ii}-based black hole mass by a factor of 1.8 ($M_\mathrm{BH}=1.6^{+0.2}_{-0.2} \times10^9 M_\odot$), as the black hole mass scales with $\mathrm{FWHM(Mg{\sc II)}}^2$ \citep{Vestergaard09} and the continuum luminosity does not change between the two models.
This difference is still within the 0.5 dex systematic uncertainty of the Mg{\sc ii}-based mass measurements \citep{Shen13_review}.
The Eddington ratio becomes $L_\mathrm{bol}/L_\mathrm{Edd}=0.60_{-0.07}^{+0.08}$ in this case.

Our updated mass measurement confirms that ULAS J1342+0928 is powered by a matured and actively accreting SMBH at the Eddington limit, which poses a question to the formation and early growth scenario of the SMBHs in the early universe \citep[e.g.,][]{Inayoshi20}.

\section{Discussion} \label{sec:discussion}
\subsection{Systematic Uncertainties on Mg{\sc ii} and Fe{\sc ii} Measurements } \label{sec:discussion_continuum}
There are offsets in the measured Fe{\sc ii}/Mg{\sc ii} flux ratios in the literature, even at the same redshift and Eddington ratio ranges.
Previous studies argued that the offsets are at least partly originated from systematic uncertainties of the Mg{\sc ii} and Fe{\sc ii} line measurements, which could be up to a factor of two and more \citep{Kurk07, DeRosa11, DeRosa14, Shin19}.
For example, SDSS J1030+0524 at $z=6.3$ \citep{Fan01} has a number of Fe{\sc ii}/Mg{\sc ii} measurements based on different observations and methods with the reported values ranging from Fe{\sc ii}/Mg{\sc ii}$=0.99$ to $8.65$ \citep{Maiolino03, Freudling03, Iwamuro04, Kurk07, Jiang07, DeRosa11}.
This section explores  how the Fe{\sc ii}/Mg{\sc ii} measurement using the deep spectrum of ULAS J1342+0928 is affected by different fitting approaches.

\begin{figure}[htb!]
\centering
 \includegraphics[width=\linewidth]{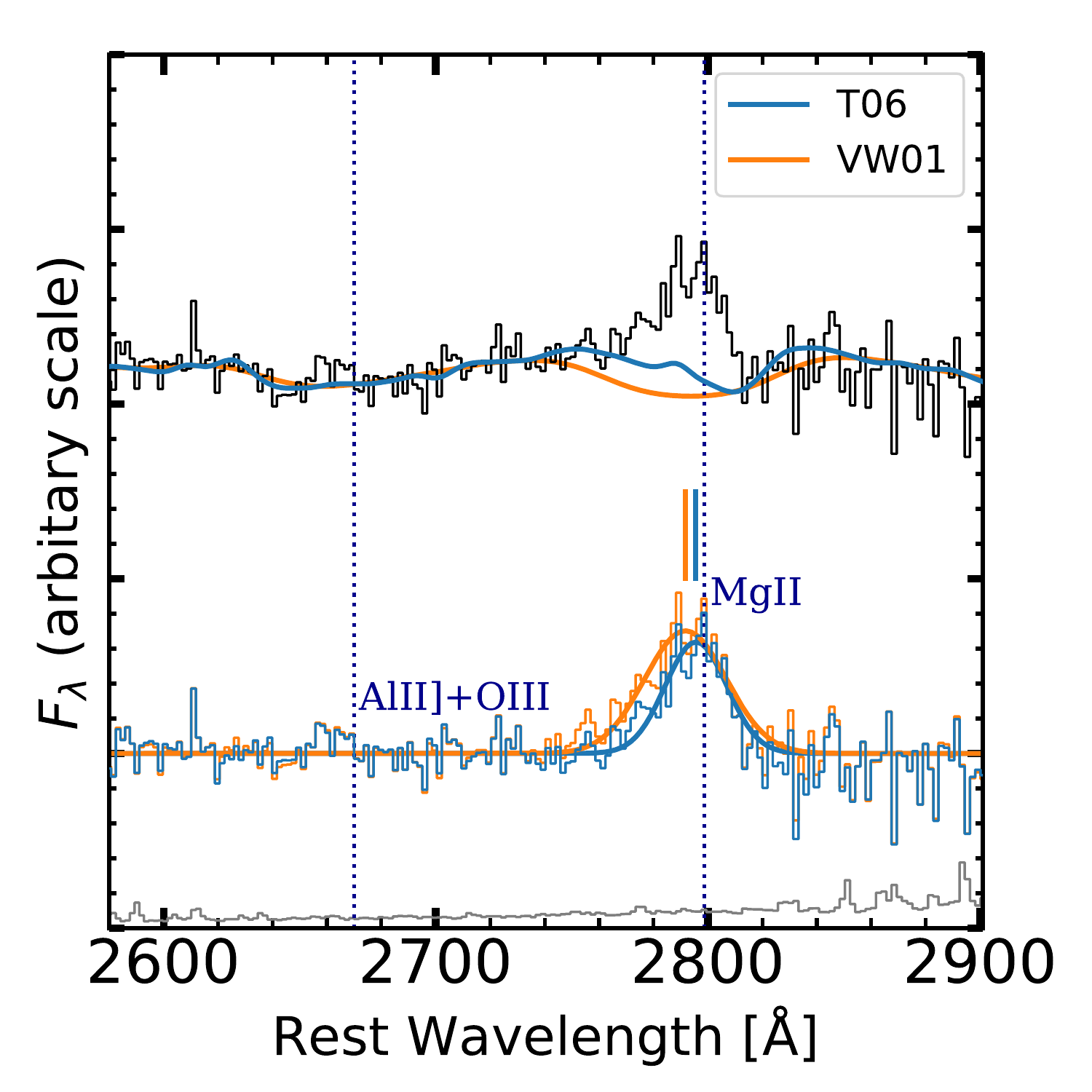}
\caption{
{\bf Top:} The continuum (power-law and Balmer) subtracted spectrum around the Mg{\sc ii} line (black).
The best-fit iron pseudo-continua from the T06 and the VW01 models are shown in blue and orange, respectively.
{\bf Bottom:}
The same spectrum after continuum and Fe{\sc ii} subtraction.
The best-fit single Gaussian profile and the line center of Mg{\sc ii} are shown for each model.
The flux excess visible at $\lambda_\mathrm{rest}\approx2670$ is from the blended and blueshifted Al{\sc ii]}+O{\sc iii}$\lambda2672$.
The rest-frame wavelengths of Mg{\sc ii} and the Al{\sc ii]}+O{\sc iii} composite are indicated by  vertical dotted lines.
The flux error at each pixel is indicated by the grey line at the bottom.
} \label{fig:MgII}
\end{figure}

\begin{figure*}[hbt!]
\centering
 \includegraphics[width=\linewidth]{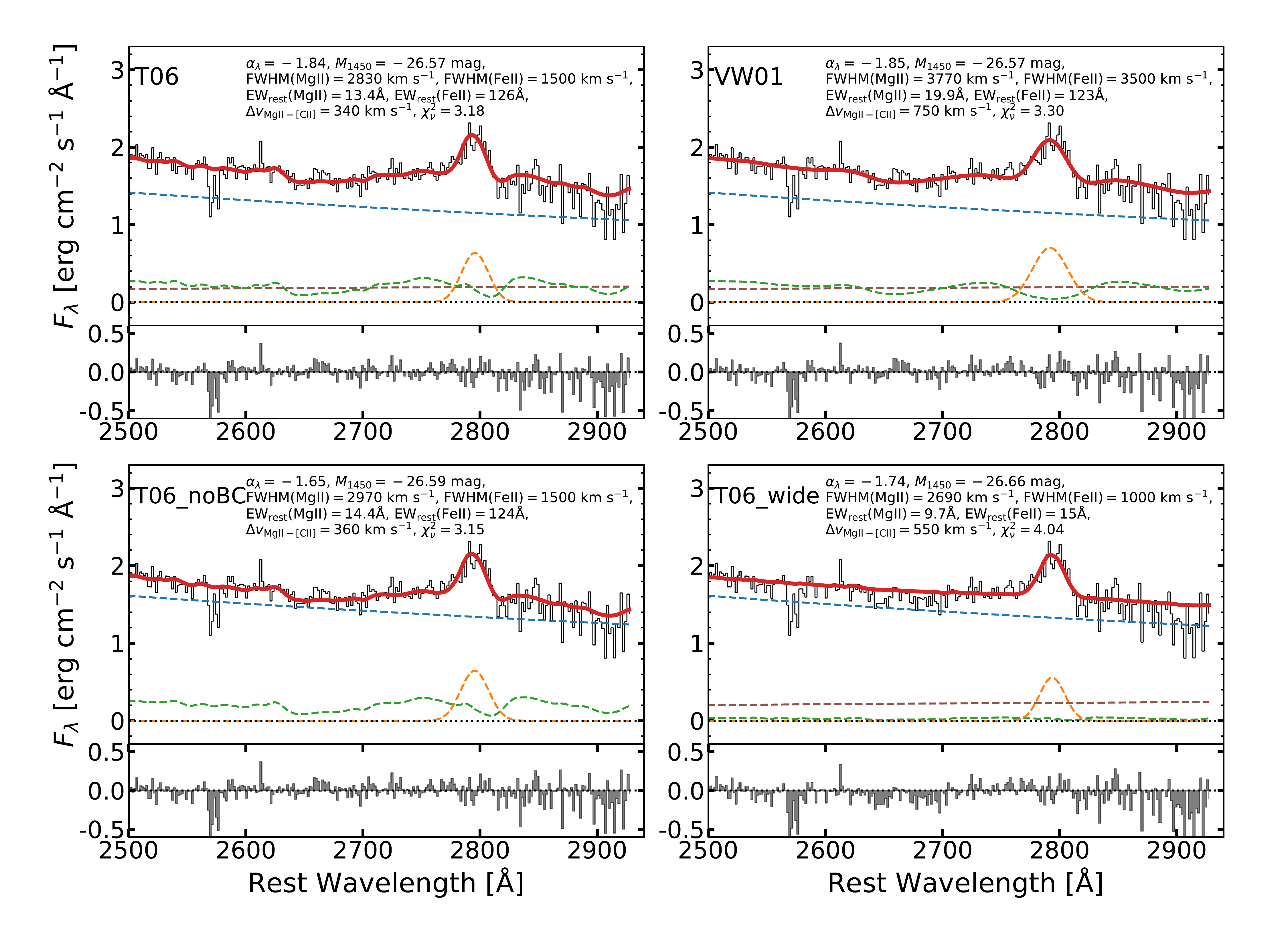}
\caption{
Comparison of four different spectral models described in Section~\ref{sec:discussion_continuum} (See Table~\ref{tab:emission_line_ratio}).
The observed spectrum is the same as the one presented in Figure~\ref{fig:spec_all}, while the residual fluxes of the fit are also shown in the bottom panel.
The name of the model is indicated at the top left of each panel.
The colors of the lines are the same as Figure~\ref{fig:spec_all}, while 
the best-fit single Gaussian for Mg{\sc ii} is also shown in orange.
In each model, the best-fit emission line and continuum parameters and reduced-chisquare of the multi-component (continuum+iron+Mg{\sc ii}) fit at $\lambda_\mathrm{rest}=$ 2500--2900\AA\  are shown inside each panel.
Note that a $z=6.86$ Mg{\sc ii} doublet absorption is at $\lambda_\mathrm{rest}=2580$\AA, which was reported in \citet{Cooper19}.
} \label{fig:MgII_comp}
\end{figure*}

First, different iron templates affect total Mg{\sc ii} flux and line profile, while this is not the case for the other emission lines at the bluer side
(Table~\ref{tab:emission_line}). 
Figure~\ref{fig:MgII} shows the continuum-subtracted spectrum at $\lambda_\mathrm{rest}=$ 2500--2900\AA, where the best-fit Fe{\sc ii} and Mg{\sc ii} components are compared between the T06 and the VW01 models.
Figure~\ref{fig:MgII_comp} also shows the two models with the decomposed spectral components (i.e., power-law continuum, Balmer continuum, iron template, and single Gaussian for the Mg{\sc ii} line) at the same wavelength range.
The difference of the power-law+Balmer continuum components at this region is tiny between the two models, compared to flux errors.
As already mentioned in Section~\ref{sec:spec_model_iron}, the difference between the two iron templates is the iron contribution underneath the Mg{\sc ii} line, where \citet{Tsuzuki06} compensated the over-subtracted Fe{\sc ii} flux with their photoionization simulation.
The T06 model results in a narrower Mg{\sc ii} line profile  and a smaller equivalent width than the VW01 model (FWHM$=2830_{-210}^{+210}$ km s$^{-1}$ and EW$_\mathrm{rest}=13.4_{-0.9}^{+0.8}$\AA\ for T06, and FWHM$=3780_{-260}^{+220}$ km s$^{-1}$ and EW$_\mathrm{rest}=19.9_{-1.5}^{+0.7}$\AA\  for VW01), while
in both cases the line width is broader than that measured in \citet[][FWHM$=2500_{-320}^{+480}$ km s$^{-1}$]{Banados18}.
The difference due to different iron templates is also found in the Mg{\sc ii} blueshift.
The residual spectra after subtracting the best-fit continuum and iron templates are compared in the bottom panel of Figure~\ref{fig:MgII}.
The Mg{\sc ii} line center in the T06 model has a smaller blueshift ($\Delta v_\mathrm{MgII-[CII]}=340_{-80}^{+110}$ km s$^{-1}$) than in the VW01 model ($\Delta v_\mathrm{MgII-[CII]}=750_{-90}^{+90}$ km s$^{-1}$).
This is because of the asymmetric profile of the iron forest underneath Mg{\sc ii} (see Fig.~\ref{fig:MgII}).
As a result, there is more remaining flux in the VW01 model at the blue side of the emission line after the iron subtraction, which enhances the Mg{\sc ii} flux by $47_{-11}^{+11}\%$ 
and reduces Fe{\sc ii}/Mg{\sc ii} (Table~\ref{tab:emission_line_ratio}).
This systematic effect on the Mg{\sc ii} measurement caused by different iron templates  was also discussed in the literature for other sources  \citep[e.g.,][]{Kurk07, Woo18, Shin19}.

Another factor that affects the emission line measurements is the fitting procedure of the quasar power-law and Balmer continuum.
As mentioned in Section~\ref{sec:spec_model}, the Balmer continuum is degenerate with the power-law continuum and the Fe{\sc ii} pseudo-continuum at the wavelength range of the GNIRS spectrum.
The Balmer continuum is tied to the power-law continuum in our spectral modeling with its scaling factor fixed to $30$\% of the power-law flux at $\lambda_\mathrm{rest}=3675$\AA.
The electron temperature and the optical depth were fixed to $T_e=15000$K and $\tau=1$, as \citet{DeRosa14} argued that directly fitting those two parameters  does not make significant impacts on the Fe{\sc ii}/Mg{\sc ii} measurements.

In order to address the robustness of those assumptions and the potential impacts on the Fe{\sc ii}/Mg{\sc ii} measurements,  different Balmer continuum models were tested with our GNIRS spectrum of ULAS J1342+0928 by changing the normalization factor to $0$--$100$\%\footnote{The scaling factors of the Balmer continuum used in this test were 0, 0.01, 0.03, 0.04, 0.05, 0.1, 0.3, 0.4, 0.5, and 1 with respect to $F^\mathrm{PL}_\lambda$ at $\lambda_\mathrm{obs}=3675(1+z)\mathrm{\AA}$. 
The 0\% model corresponds to our T06\_noBC model.}.
The electron temperature and optical depth were fixed to the same values as in the original model.
Different values of the two parameters were also tested, but their impacts on the emission line measurements are smaller than those of the normalization, confirming the test done by \citet{DeRosa14}.
The broad emission lines were fitted with the same procedure applied to our fiducial model with the T06 iron template.
As a result, the GNIRS spectrum was well fitted when the normalization factors were $\leq30$\%, as the goodness of the joint continuum and iron fitting at the Fe{\sc ii} windows becomes slightly better ($-0.02 \lesssim \Delta\chi^2_\nu\lesssim0$).
The reduced chisquare got worse when the normalization factor was higher ($>30$\%) up to $\Delta\chi^2_\nu=0.3$, as the Fe{\sc ii} forest was poorly fitted due to the stronger-than-necessary Balmer continuum at $\lambda_\mathrm{rest}\gtrsim2200$\AA.
Therefore, it is likely that the actual contribution of the Balmer continuum is $30$\% or lower, while the exact amount is beyond the scope of this paper.

The resulting range of the Fe{\sc ii}/Mg{\sc ii} values is $7.8\leq$ Fe{\sc ii}/Mg{\sc ii} $\leq8.7$ with the minimum value obtained when the normalization is 3--5\%.
Those models actually give the minimum chisquare in the continuum+iron fitting, while the 
deviation from the fiducial 30\% model is within the $1\sigma$ uncertainty (Fe{\sc ii}/Mg{\sc ii}$=8.7_{-1.3}^{+0.6}$).
Therefore, the 0--30\% normalization gives adequate spectral decomposition and a reliable measurement of Fe{\sc ii}/Mg{\sc ii} for ULAS J1342+0928.
In Table~\ref{tab:emission_line_ratio}, the flux ratios of the power-law-only continuum model dubbed as ``T06\_noBC" are reported to represent the ``weak Balmer continuum" models.
The best-fit spectral decomposition in the T06\_noBC model is also shown in Figure~\ref{fig:MgII_comp}.
In this case, the power-law slope gets flatter ($\alpha_\lambda=-1.65_{-0.01}^{+0.02}$) with little change in the absolute magnitude ($M_{1450}=-26.59\pm0.04$).
The Fe{\sc ii}/Mg{\sc ii} line ratio decreases by 11\% (Fe{\sc ii}/Mg{\sc ii}$=7.7_{-0.9}^{+0.6}$), which is within the $1\sigma$ uncertainty of the T06 model.
This small change indicates that the assumption on the Balmer continuum strength is a relatively small factor that introduces systematic uncertainties in the Fe{\sc ii} and Mg{\sc ii} flux measurements.

On the other hand, the continuum window used in fitting the power-law continuum has a bigger impact on the Mg{\sc ii} and Fe{\sc ii} measurements.
In our original T06 modeling, the following emission line-free regions were chosen: 
$\lambda_\mathrm{rest}=$ 1275--1285\AA, 1310--1325\AA, 2500--2750\AA, and 2850--2890\AA\ (Section~\ref{sec:spec_model_fit}).
Only a small region of the quasar continuum is used at the blue side because most parts of the observed wavelengths are covered with the extremely broad emission lines of ULAS J1342+0928 (FWHM $>10,000$ km s$^{-1}$ for high-ionization lines;  Table~\ref{tab:emission_line}).
The same spectral fitting was performed with a wider continuum window to test how the Fe{\sc ii}/Mg{\sc ii} line ratio is sensitive to the continuum fitting (``T06\_wide" model in Table~\ref{tab:emission_line_ratio}).
Three additional wavelength ranges, $\lambda_\mathrm{rest}=$ 1425--1470\AA, 1680--1710\AA, and 1975--2050\AA\ were added to the original continuum window.
Also, the second reddest window was extended from 2500--2750\AA\ to 2260--2750\AA\ to cover the entire UV Fe{\sc ii} bump.
Those regions are in fact not line-free.
The observed flux deviates from the power-law component at $\lambda_\mathrm{rest}\approx$ 1600--1800\AA\  in the composite spectra of low-redshift quasars \citep{VB01, Nagao06b}.
The rest-frame $\lambda_\mathrm{rest}=$ 1425--1470\AA\ is usually a line-free region, but the extremely broad outskirts of the Si{\sc iv} and C{\sc iv} emission lines in ULAS J1342+0928 likely contribute to the observed flux in this region (see Figure~\ref{fig:spec_all}).
There are Fe{\sc iii} emission lines at $\lambda_\mathrm{rest}\sim$ 2000\AA.

As a result, the power-law continuum becomes brighter than the T06 model by $\Delta M_{1450}=-0.09$ mag with a flatter slope of $\alpha_\lambda=-1.74_{-0.01}^{+0.01}$.
More importantly, the higher continuum level largely suppresses the iron contribution  (Figure~\ref{fig:MgII_comp}).
The total Fe{\sc ii} flux becomes only 14\% of the one in the T06 model, which results in a smaller line ratio of Fe{\sc ii}/Mg{\sc ii}$=1.4_{-0.6}^{+0.4}$.
This model better traces the observed flux at $\lambda_\mathrm{rest}\approx 1700$\AA\ and $\approx 2000$\AA, while parts of the flux in those regions are from the unidentified weak emission lines. 
This model poorly traces the Fe{\sc ii} bump compared to the other three models.
The reduced-chisquare of the continuum+iron+Mg{\sc ii} line fitting at $\lambda_\mathrm{rest}=$ 2500--2900\AA\ gets worse by $\Delta \chi_\nu=0.86$ from the T06 model.
The T06 model (and also the VW01 and the T06\_noBC models) reproduces the UV Fe{\sc ii} bump, while the T06\_wide model roughly fits this feature only with the power-law continuum.
This overestimated continuum level of the T06\_wide model also affects the other emission lines at bluer sides, while the effect is smaller than that for Fe{\sc ii}/Mg{\sc ii}.
As reported in Table~\ref{tab:emission_line_ratio}, the emission line ratios other than Fe{\sc ii}/Mg{\sc ii} also decrease from the T06 model by $\leq11$\%.

Overall,  good care should be taken in the continuum and iron modeling, especially when  Fe{\sc ii}/Mg{\sc ii} is to be measured from a quasar spectrum.
Although not addressed in this section, there are other factors that potentially affect the Fe{\sc ii}/Mg{\sc ii} measurements, such as whether or not the blueshifts of the iron emission lines are considered.
Previous studies reported large Fe{\sc ii}/Mg{\sc ii} scatters even at the same redshift and luminosity ranges, with some quasars showing very small Fe{\sc ii}/Mg{\sc ii} ratios \citep[Fe{\sc ii}/Mg{\sc ii}$\ll1$;][]{Mazzucchelli17, Shin19}.
Those results may indicate that there is a variation in the iron abundance and the iron enrichment is ongoing in some quasars at $z\lesssim7$.
However, given the large effects that different spectral modeling approaches have on the Fe{\sc ii}/Mg{\sc ii} measurements as presented in this section,
it is important to revisit the early iron enrichment by measuring Fe{\sc ii}/Mg{\sc ii} for a large sample of quasars spanning over a wide redshift range based on a unified approach.

\subsection{Early Iron Enrichment } \label{sec:discussion_Fe}
The BLR gas-phase metallicity traces the star-formation history of the host galaxies of quasars, as the BLR gas originated from the ISM of the host galaxies falling onto the nuclear regions.
Previous studies show that the BLR clouds have supersolar metallicity up to $z\sim7$ \citep[e.g.,][]{Nagao06b, Jiang07, DeRosa14, Xu18, Tang19}.
Moreover, the lack of redshift evolution in the Fe{\sc ii}/Mg{\sc ii} line ratios (with a large scatter) indicates rapid iron enrichment with respect to the $\alpha$-elements within the first billion years of the universe \citep[e.g.,][]{Jiang07, DeRosa14, Mazzucchelli17}.
This trend holds for ULAS J1342+0928 at $z=7.54$ (Section~\ref{sec:spec_result_line_ratio} and Section~\ref{sec:iron_abundance}).

While SNe Ia have a higher iron production efficiency than SNe II, 
the SNe Ia explosion are delayed from the initial starburst.
This time delay is due to the timescale required for formation of white dwarfs, and mass transfer from their companion stars or mergers due to the loss of orbital energy by gravitational wave radiation \citep{Maoz14}.
For a star with a mass a few times solar to explode as a SN Ia, the time delay is $\sim1$ Gyr \citep{Greggio83}, while the minimum time delay is $\sim40$ Myr, which corresponds to the main-sequence lifetime of an $8M_\odot$ star.

The recent observational constraints on the SN Ia rate as a function of times between starbursts and explosion 
(or so-called delay-time distribution) 
favor a power-law shape with a slope of $t^{-1}$ over $t\approx$ 0.1--10 Gyr \citep[e.g.,][and references therein]{Maoz12, Maoz14}\footnote{
According to the SN Ia rate given in \citet[][Eq.~13]{Maoz12}, the number of SN Ia between $40$ Myr and $680$ Myr is $10^7$ SNe per $10^{10}M_\odot$ host stellar mass.}.
\citet{Rodney14} suggest that about half of the SNe Ia explode within $500$ Myr.
Therefore, given the young age of the universe when ULAS J1342+0928 was observed ($680$ Myr), it is likely that those prompt SNe Ia contributed to the iron enrichment of the host ISM.

It is useful to estimate approximately how many SNe Ia and II are needed to explain the early iron enrichment in the BLR gas.
Here, the BLR gas clouds are assumed to be $10^4M_\odot$ \citep{Baldwin03} with $5Z_\odot$ metallicity \citep{Nagao06b}.
If the relative mass fraction of metal elements is the same as the solar system \citep{Asplund09}, the magnesium and iron masses in the BLR are $36M_\odot$ and $65M_\odot$, respectively.
Based on the SNe Ia yields of \citet{Iwamoto99}\footnote{The SNe Ia yields quoted here are based on their W7 model.} and SNe II yields of \citet{Nomoto13},
those masses can be achieved with $\sim60$ SNe Ia and $\sim200$ SNe II of $20M_\odot$ stars with zero-to-super solar metallicity ($Z=0$--$0.05$). 
Pair-instability supernovae (PISNe) could also significantly contribute to the iron enrichment if the Pop-III stars are as massive as $140<M/M_\odot<260$.
$7M_\odot$ of iron is ejected from a metal-free $200M_\odot$ star  through PISNe in this case \citep{Nomoto13}.

However, more SNe should have been responsible for the high BLR metallicity than the numbers given above.
The chemical enrichment of the BLR gas is tied to the host star formation at a much larger scale than the actual BLR size, because the BLR gas is likely originated from the host ISM inflowing onto the central SMBH.
A non-negligible fraction of the metal-polluted SNe remnants are consumed in the subsequent star formation and thus do not remain as gas.
Moreover, the BLR gas with $10^{3-4}M_\odot$ only accounts for a tiny fraction of the mass at the nuclear region, given the central SMBH mass of $M_\mathrm{BH}\sim10^{9-10}M_\odot$ for a luminous quasar as those known at $z>6$.
Therefore, it is more realistic that the high BLR metallicity reflects far more than $10$--$100$ SNe, and
as \citet{Baldwin03} concluded, the rapid chemical enrichment of the BLR gas should trace the star-formation history of the quasar host galaxies at the scale where the local ISM is tied to the SMBH feeding.

\section{Summary and Conclusion} \label{sec:summmary}
\subsection{Summary of This Work} \label{sec:summmary_work}
The BLR properties of ULAS J1342+0928 at $z=7.54$ were measured with a deep (9 hours on source) NIR spectrum taken by Gemini/GNIRS at $\lambda_\mathrm{rest}=$ 970--2930\AA.
The spectrum was modeled by a combination of a power-law continuum, a Balmer continuum, and templates of UV iron (Fe{\sc ii}+Fe{\sc iii}) pseudo-continuum.
Various emission lines are then modeled by single Gaussian profiles after subtracting those continuum components, namely Si{\sc ii}$\lambda1263$, Si{\sc iv}$\lambda1397$, C{\sc iv}$\lambda1549$, Al{\sc iii}$\lambda 1857$, C{\sc iii]}$\lambda1909$, and Mg{\sc ii}$\lambda2798$.
The line flux of Ly$\alpha$+N{\sc v}$\lambda 1240$ composite was instead measured by summing the flux above the continuum at $\lambda_\mathrm{rest}=$ 1160--1290\AA.
N{\sc v}$\lambda1240$ and He{\sc ii}$\lambda1640$ were not identified as individual lines due to their heavy line blending, weakness, atmospheric absorption, and the strong IGM absorption (for N{\sc v}).

ULAS J1342+0928 exhibits large blueshifts with respect to its systematic redshift from the [C{\sc ii}] $158\micron$ line.
The amount of blueshifts has a linear correlation with the ionization potentials up to $\Delta v_\mathrm{[CII]}=5510_{-110}^{+240}$ km s$^{-1}$ that is observed for C{\sc iv}.
The high-ionization emission lines also show broader profiles than those of low-ionization emission lines with FWHM more than 10,000 km s$^{-1}$.
Those velocity offsets are remarkably larger than the average of $z\sim$ 6--7 quasars compiled in \citet{Shen19} and \citet{Meyer19}, suggesting extreme outflow components in the BLR clouds.

The virial BH mass measurement was revised from \citet{Banados18} based on the multiple-component fit in this work.
The Mg{\sc ii}-based black hole mass based on our T06 model is $M_\mathrm{BH}=9.1_{-1.4}^{+1.3} \times 10^8 M_\odot$ with the Eddington ratio of $L_\mathrm{bol}/L_\mathrm{Edd}=1.1_{-0.2}^{+0.2}$.
This result confirms that ULAS J1342+0928 is powered by an already matured SMBH actively accreting at the near-Eddington accretion rate.

The measured emission line ratios, namely Si{\sc ii}/C{\sc iv}, Si{\sc iv}/C{\sc iv}, Al{\sc iii}/C{\sc iv}, and C{\sc iii]}/C{\sc iv}, suggest that the BLR gas of ULAS J1342+0928 had a super-solar metallicty when the age of the universe was only $680$ Myr.
While the BLR gas-phase metallicity traces the past star-formation at the galaxy center,
this picture is consistent with the dust-rich and near-solar metallicity ISM of the host galaxy revealed by the rest-FIR observations of the host galaxy \citep{Venemans17, Novak19}.

The Fe{\sc ii}/Mg{\sc ii} emission line ratio, a tracer of iron enrichment in the early universe, is also comparable to those of the lower-redshift quasars (Fe{\sc ii}/Mg{\sc ii} $=8.7_{-1.3}^{+0.6}$).
The Fe{\sc ii}/Mg{\sc ii} becomes even more typical when its Eddington ratio dependence is taken into account \citep{Sameshima17}.
Our result is somewhat in conflict with the predictions of SNe Ia nucleosynthesis, the timescale of which is $\sim1$ Gyr.
Prompt SNe Ia (and perhaps PISNe) that exploded within $<680$ Myr from the first star formation would have contributed to the early iron enrichment at the galaxy center.
The non-evolution trend seen in the BLR line ratios including Fe{\sc ii}/Mg{\sc ii} should reflect the early metal enrichment at the much wider scale than the BLR itself, where the local ISM of the host galaxy is tied to the SMBH feeding.

There are a number of factors that introduce systematic errors in the Fe{\sc ii}/Mg{\sc ii} measurements.
Four models with different iron templates (empirical templates from \citealt{Tsuzuki06} and \citealt{Vestergaard01}), continuum components (power-law plus Balmer continuum, or power-law only), and wavelength ranges at which the spectrum is fitted  were tested to
investigate how much the Fe{\sc ii}/Mg{\sc ii} value of ULAS J1342+0928 is sensitive to the fitting methods.
As a result, not only the choice of iron templates, which has been discussed in the literature, but also the continuum window has a significant impact on Fe{\sc ii}/Mg{\sc ii}.
The contribution of the quasar continuum is overestimated if it is fitted at the wavelength regions where weak and individually unidentified emission lines are present.
The Fe{\sc ii} flux is underestimated in this case, which potentially explains the relatively low Fe{\sc ii}/Mg{\sc ii} ratios reported in the literature at high redshift.
Given the large offsets that different fitting approaches introduce, 
one should measure the Fe{\sc ii}/Mg{\sc ii} line ratios with a unified approach over a wide redshift range to address the potential fluctuation of the iron abundance in the BLR clouds.
The Eddington ratio dependence of Fe{\sc ii}/Mg{\sc ii} should also be corrected to translate the line ratio to the actual abundance ratio, as introduced by \citet{Sameshima17}.

\subsection{Future Prospects} \label{sec:summmary_future}
There is apparently no redshift evolution in the BLR properties up to $z=7.54$, except the BLR blueshifts (Section~\ref{sec:spec_result_line}).
The BLR studies of even higher-redshift (i.e., $z\gg8$) quasars are required to identify the epoch when the BLR clouds were metal polluted to the super-solar metallicity.
One caveat in the previous and the present studies is that most of the studies at high redshift are biased toward luminous quasars at each epoch.
There is an indirect correlation between the UV luminosity of quasars and their host metallicity \citep[e.g.,][]{Matsuoka11}.
Those two quantities are tied through the relation between SMBH and host bulge mass, and the mass-metallicity relation of star-forming galaxies.
In other words, studies on luminous quasars selectively sample massive SMBHs, the host galaxies of which are also presumably massive.
The host galaxies of $z\gtrsim6$ luminous quasars have dynamical mass of $M_\mathrm{dyn}\sim10^{11}M_\odot$ \citep[e.g.,][]{Decarli18}.
Such matured galaxies would have already experienced chemical enrichment through their past star formation compared to lower-mass galaxies at the same redshift.

In this sense, lower-luminosity quasars could be better targets to trace chemically young BLRs.
\citet{Shin19} was motivated by this expectation and measured the Fe{\sc ii}/Mg{\sc ii} ratios for $z\sim3$ quasars that are
an order of magnitude fainter ($L_\mathrm{bol}\sim10^{46.5}$ erg s$^{-1}$) than the ones in previous studies at the same redshift; however they did not find any significant difference in Fe{\sc ii}/Mg{\sc ii}.
At $z\gtrsim6$, the attempts of finding low-luminosity quasars have been led by the optical wide-field survey of the Subaru/Hyper Suprime-Cam \citep[e.g.,][]{Matsuoka16, Matsuoka19a}.
This deep survey has revealed SMBHs at less massive range \citep[$M_\mathrm{BH}=10^{7-9}M_\odot$,][]{Onoue19}.
The host galaxies of those HSC quasars also have a variety of dynamical mass down to  $M_\mathrm{dyn}\sim10^{10}M_\odot$ \citep{Izumi18, Izumi19}.
Therefore, the deep observations of those less extreme SMBH populations at $z>6$ have a potential to witness metal-poor BLRs, while the sensitivity of $30$m-class ground-based telescopes or next-generation large space telescopes are needed to detect weak emission lines from the faintest $z=6$--$7$ HSC quasars at $M_{1450}\sim-22$ mag (the continuum level of which is $\sim10^{-19}$ erg s$^{-1}$ cm$^{-2}$ \AA$^{-1}$ in the observed frame).

Also interestingly, recent observations have identified a few luminous $z\sim6$ quasars which were observed possibly only after $<10^5$ years from their ignition \citep{Eilers17}.
While their SMBHs are as massive as those of the other luminous quasars at the same redshift range \citep[$M_\mathrm{BH}\sim10^9M_\odot$;][]{Eilers18},
the young quasars could also be good targets for future observations to identify the less metal-enriched BLRs, if their host galaxies are also young and host metal-poor ISMs.

\acknowledgments
We are greateful to the referee for providing constructive comments on the manuscript.
We thank the Gemini North staffs to execute our programs.
We also thank H. Sameshima for providing us his data on the SDSS DR7 quasars.
This work was supported by the ERC Advanced Grant 740246 ``Cosmic gas".
F. Wang thanks the support provided by NASA through the NASA Hubble Fellowship grant \#HST-HF2-51448.001-A awarded by the Space Telescope Science Institute, which is operated by the Association of Universities for Research in Astronomy, Incorporated, under NASA contract NAS5-26555.

The data presented in this paper is based on observations obtained at the Gemini Observatory (GN-2017A-DD-4, GN-2019A-FT-115).
The Gemini Observatory is operated by the Association of Universities for Research in Astronomy, Inc., under a cooperative agreement with the NSF on behalf of the Gemini partnership: the National Science Foundation (United States), National Research Council (Canada), CONICYT (Chile), Ministerio de Ciencia, Tecnolog\'{i}a e Innovaci\'{o}n Productiva (Argentina), Minist\'{e}rio da Ci\^{e}ncia, Tecnologia e Inova\c{c}\~{a}o (Brazil), and Korea Astronomy and Space Science Institute (Republic of Korea).

\vspace{5mm}
\facilities{Gemini-North (GNIRS)}

\software{astropy \citep{Astropy},  
    }

\bibliographystyle{aasjournal}
\bibliography{ref_Pisco_NIR1}

\begin{thebibliography}{}
\expandafter\ifx\csname natexlab\endcsname\relax\def\natexlab#1{#1}\fi
\providecommand{\url}[1]{\href{#1}{#1}}

\bibitem[{{Asplund} {et~al.}(2009){Asplund}, {Grevesse}, {Sauval}, \&
  {Scott}}]{Asplund09}
{Asplund}, M., {Grevesse}, N., {Sauval}, A.~J., \& {Scott}, P. 2009, \araa, 47,
  481

\bibitem[{{Astropy Collaboration} {et~al.}(2013){Astropy Collaboration},
  {Robitaille}, {Tollerud}, {Greenfield}, {Droettboom}, {Bray}, {Aldcroft},
  {Davis}, {Ginsburg}, {Price-Whelan}, {Kerzendorf}, {Conley}, {Crighton},
  {Barbary}, {Muna}, {Ferguson}, {Grollier}, {Parikh}, {Nair}, {Unther},
  {Deil}, {Woillez}, {Conseil}, {Kramer}, {Turner}, {Singer}, {Fox}, {Weaver},
  {Zabalza}, {Edwards}, {Azalee Bostroem}, {Burke}, {Casey}, {Crawford},
  {Dencheva}, {Ely}, {Jenness}, {Labrie}, {Lim}, {Pierfederici}, {Pontzen},
  {Ptak}, {Refsdal}, {Servillat}, \& {Streicher}}]{Astropy}
{Astropy Collaboration}, {Robitaille}, T.~P., {Tollerud}, E.~J., {et~al.} 2013,
  \aap, 558, A33

\bibitem[{{Ba{\~n}ados} {et~al.}(2015){Ba{\~n}ados}, {Decarli}, {Walter},
  {Venemans}, {Farina}, \& {Fan}}]{Banados15b}
{Ba{\~n}ados}, E., {Decarli}, R., {Walter}, F., {et~al.} 2015, ApJL, 805, L8

\bibitem[{{Ba{\~n}ados} {et~al.}(2016){Ba{\~n}ados}, {Venemans}, {Decarli},
  {Farina}, {Mazzucchelli}, {Walter}, {Fan}, {Stern}, {Schlafly}, {Chambers},
  {Rix}, {Jiang}, {McGreer}, {Simcoe}, {Wang}, {Yang}, {Morganson}, {De Rosa},
  {Greiner}, {Balokovi{\'c}}, {Burgett}, {Cooper}, {Draper}, {Flewelling},
  {Hodapp}, {Jun}, {Kaiser}, {Kudritzki}, {Magnier}, {Metcalfe}, {Miller},
  {Schindler}, {Tonry}, {Wainscoat}, {Waters}, \& {Yang}}]{Banados16}
{Ba{\~n}ados}, E., {Venemans}, B.~P., {Decarli}, R., {et~al.} 2016, \apjs, 227,
  11

\bibitem[{{Ba{\~n}ados} {et~al.}(2018){Ba{\~n}ados}, {Venemans},
  {Mazzucchelli}, {Farina}, {Walter}, {Wang}, {Decarli}, {Stern}, {Fan},
  {Davies}, {Hennawi}, {Simcoe}, {Turner}, {Rix}, {Yang}, {Kelson}, {Rudie}, \&
  {Winters}}]{Banados18}
{Ba{\~n}ados}, E., {Venemans}, B.~P., {Mazzucchelli}, C., {et~al.} 2018, \nat,
  553, 473

\bibitem[{{Ba{\~n}ados} {et~al.}(2019){Ba{\~n}ados}, {Novak}, {Neeleman},
  {Walter}, {Decarli}, {Venemans}, {Mazzucchelli}, {Carilli}, {Wang}, {Fan},
  {Farina}, \& {Rix}}]{Banados19}
{Ba{\~n}ados}, E., {Novak}, M., {Neeleman}, M., {et~al.} 2019, \apjl, 881, L23

\bibitem[{{Baldwin} {et~al.}(2003){Baldwin}, {Ferland}, {Korista}, {Hamann}, \&
  {Dietrich}}]{Baldwin03}
{Baldwin}, J.~A., {Ferland}, G.~J., {Korista}, K.~T., {Hamann}, F., \&
  {Dietrich}, M. 2003, \apj, 582, 590

\bibitem[{{Baldwin} {et~al.}(2004){Baldwin}, {Ferland}, {Korista}, {Hamann}, \&
  {LaCluyz{\'e}}}]{Baldwin04}
{Baldwin}, J.~A., {Ferland}, G.~J., {Korista}, K.~T., {Hamann}, F., \&
  {LaCluyz{\'e}}, A. 2004, \apj, 615, 610

\bibitem[{{Barth} {et~al.}(2003){Barth}, {Martini}, {Nelson}, \&
  {Ho}}]{Barth03}
{Barth}, A.~J., {Martini}, P., {Nelson}, C.~H., \& {Ho}, L.~C. 2003, \apjl,
  594, L95

\bibitem[{{Boroson}(2005)}]{Boroson05}
{Boroson}, T. 2005, \aj, 130, 381

\bibitem[{{Carilli} {et~al.}(2007){Carilli}, {Neri}, {Wang}, {Cox}, {Bertoldi},
  {Walter}, {Fan}, {Menten}, {Wagg}, {Maiolino}, {Omont}, {Strauss},
  {Riechers}, {Lo}, {Bolatto}, \& {Scoville}}]{Carilli07}
{Carilli}, C.~L., {Neri}, R., {Wang}, R., {et~al.} 2007, ApJL, 666, L9

\bibitem[{{Clough} {et~al.}(2005){Clough}, {Shephard}, {Mlawer}, {Delamere},
  {Iacono}, {Cady-Pereira}, {Boukabara}, \& {Brown}}]{Clough05}
{Clough}, S.~A., {Shephard}, M.~W., {Mlawer}, E.~J., {et~al.} 2005, \jqsrt, 91,
  233

\bibitem[{{Coatman} {et~al.}(2017){Coatman}, {Hewett}, {Banerji}, {Richards},
  {Hennawi}, \& {Prochaska}}]{Coatman17}
{Coatman}, L., {Hewett}, P.~C., {Banerji}, M., {et~al.} 2017, \mnras, 465, 2120

\bibitem[{{Cooper} {et~al.}(2019){Cooper}, {Simcoe}, {Cooksey}, {Bordoloi},
  {Miller}, {Furesz}, {Turner}, \& {Ba{\~n}ados}}]{Cooper19}
{Cooper}, T.~J., {Simcoe}, R.~A., {Cooksey}, K.~L., {et~al.} 2019, \apj, 882,
  77

\bibitem[{{Davies} {et~al.}(2018){Davies}, {Hennawi}, {Ba{\~n}ados},
  {Luki{\'c}}, {Decarli}, {Fan}, {Farina}, {Mazzucchelli}, {Rix}, {Venemans},
  {Walter}, {Wang}, \& {Yang}}]{Davies18}
{Davies}, F.~B., {Hennawi}, J.~F., {Ba{\~n}ados}, E., {et~al.} 2018, \apj, 864,
  142

\bibitem[{{De Rosa} {et~al.}(2011){De Rosa}, {Decarli}, {Walter}, {Fan},
  {Jiang}, {Kurk}, {Pasquali}, \& {Rix}}]{DeRosa11}
{De Rosa}, G., {Decarli}, R., {Walter}, F., {et~al.} 2011, \apj, 739, 56

\bibitem[{{De Rosa} {et~al.}(2014){De Rosa}, {Venemans}, {Decarli}, {Gennaro},
  {Simcoe}, {Dietrich}, {Peterson}, {Walter}, {Frank}, {McMahon}, {Hewett},
  {Mortlock}, \& {Simpson}}]{DeRosa14}
{De Rosa}, G., {Venemans}, B.~P., {Decarli}, R., {et~al.} 2014, \apj, 790, 145

\bibitem[{{Decarli} {et~al.}(2018){Decarli}, {Walter}, {Venemans},
  {Ba{\~n}ados}, {Bertoldi}, {Carilli}, {Fan}, {Farina}, {Mazzucchelli},
  {Riechers}, {Rix}, {Strauss}, {Wang}, \& {Yang}}]{Decarli18}
{Decarli}, R., {Walter}, F., {Venemans}, B.~P., {et~al.} 2018, \apj, 854, 97

\bibitem[{{Diamond-Stanic} {et~al.}(2009){Diamond-Stanic}, {Fan}, {Brandt},
  {Shemmer}, {Strauss}, {Anderson}, {Carilli}, {Gibson}, {Jiang}, {Kim},
  {Richards}, {Schmidt}, {Schneider}, {Shen}, {Smith}, {Vestergaard}, \&
  {Young}}]{Diamond-Stanic09}
{Diamond-Stanic}, A.~M., {Fan}, X., {Brandt}, W.~N., {et~al.} 2009, \apj, 699,
  782

\bibitem[{{Dietrich} {et~al.}(2002{\natexlab{a}}){Dietrich}, {Appenzeller},
  {Vestergaard}, \& {Wagner}}]{Dietrich02}
{Dietrich}, M., {Appenzeller}, I., {Vestergaard}, M., \& {Wagner}, S.~J.
  2002{\natexlab{a}}, \apj, 564, 581

\bibitem[{{Dietrich} {et~al.}(2003{\natexlab{a}}){Dietrich}, {Hamann},
  {Appenzeller}, \& {Vestergaard}}]{Dietrich03}
{Dietrich}, M., {Hamann}, F., {Appenzeller}, I., \& {Vestergaard}, M.
  2003{\natexlab{a}}, \apj, 596, 817

\bibitem[{{Dietrich} {et~al.}(2003{\natexlab{b}}){Dietrich}, {Hamann},
  {Shields}, {Constantin}, {Heidt}, {J{\"a}ger}, {Vestergaard}, \&
  {Wagner}}]{Dietrich03a}
{Dietrich}, M., {Hamann}, F., {Shields}, J.~C., {et~al.} 2003{\natexlab{b}},
  \apj, 589, 722

\bibitem[{{Dietrich} {et~al.}(2002{\natexlab{b}}){Dietrich}, {Hamann},
  {Shields}, {Constantin}, {Vestergaard}, {Chaffee}, {Foltz}, \&
  {Junkkarinen}}]{Dietrich02b}
---. 2002{\natexlab{b}}, \apj, 581, 912

\bibitem[{{Dong} {et~al.}(2011){Dong}, {Wang}, {Ho}, {Wang}, {Fan}, {Wang},
  {Zhou}, \& {Yuan}}]{Dong11}
{Dong}, X.-B., {Wang}, J.-G., {Ho}, L.~C., {et~al.} 2011, \apj, 736, 86

\bibitem[{{Eilers} {et~al.}(2017){Eilers}, {Davies}, {Hennawi}, {Prochaska},
  {Luki{\'c}}, \& {Mazzucchelli}}]{Eilers17}
{Eilers}, A.-C., {Davies}, F.~B., {Hennawi}, J.~F., {et~al.} 2017, \apj, 840,
  24

\bibitem[{{Eilers} {et~al.}(2018){Eilers}, {Hennawi}, \& {Davies}}]{Eilers18}
{Eilers}, A.-C., {Hennawi}, J.~F., \& {Davies}, F.~B. 2018, \apj, 867, 30

\bibitem[{{Fan} {et~al.}(2001){Fan}, {Narayanan}, {Lupton}, {Strauss}, {Knapp},
  {Becker}, {White}, {Pentericci}, {Leggett}, {Haiman}, {Gunn}, {Ivezi{\'c}},
  {Schneider}, {Anderson}, {Brinkmann}, {Bahcall}, {Connolly}, {Csabai}, {Doi},
  {Fukugita}, {Geballe}, {Grebel}, {Harbeck}, {Hennessy}, {Lamb}, {Miknaitis},
  {Munn}, {Nichol}, {Okamura}, {Pier}, {Prada}, {Richards}, {Szalay}, \&
  {York}}]{Fan01}
{Fan}, X., {Narayanan}, V.~K., {Lupton}, R.~H., {et~al.} 2001, \aj, 122, 2833

\bibitem[{{Fan} {et~al.}(2006){Fan}, {Strauss}, {Becker}, {White}, {Gunn},
  {Knapp}, {Richards}, {Schneider}, {Brinkmann}, \& {Fukugita}}]{Fan06}
{Fan}, X., {Strauss}, M.~A., {Becker}, R.~H., {et~al.} 2006, \aj, 132, 117

\bibitem[{{Freudling} {et~al.}(2003){Freudling}, {Corbin}, \&
  {Korista}}]{Freudling03}
{Freudling}, W., {Corbin}, M.~R., \& {Korista}, K.~T. 2003, \apjl, 587, L67

\bibitem[{{Ge} {et~al.}(2019){Ge}, {Zhao}, {Bian}, \& {Frederick}}]{Ge19}
{Ge}, X., {Zhao}, B.-X., {Bian}, W.-H., \& {Frederick}, G.~R. 2019, \aj, 157,
  148

\bibitem[{{Grandi}(1982)}]{Grandi82}
{Grandi}, S.~A. 1982, \apj, 255, 25

\bibitem[{{Gravity Collaboration} {et~al.}(2018){Gravity Collaboration},
  {Sturm}, {Dexter}, {Pfuhl}, {Stock}, {Davies}, {Lutz}, {Cl{\'e}net},
  {Eckart}, {Eisenhauer}, {Genzel}, {Gratadour}, {H{\"o}nig}, {Kishimoto},
  {Lacour}, {Millour}, {Netzer}, {Perrin}, {Peterson}, {Petrucci}, {Rouan},
  {Waisberg}, {Woillez}, {Amorim}, {Brandner}, {F{\"o}rster Schreiber},
  {Garcia}, {Gillessen}, {Ott}, {Paumard}, {Perraut}, {Scheithauer},
  {Straubmeier}, {Tacconi}, \& {Widmann}}]{GRAVITY18}
{Gravity Collaboration}, {Sturm}, E., {Dexter}, J., {et~al.} 2018, \nat, 563,
  657

\bibitem[{{Greene} {et~al.}(2019){Greene}, {Strader}, \& {Ho}}]{Greene20}
{Greene}, J.~E., {Strader}, J., \& {Ho}, L.~C. 2019, arXiv e-prints,
  arXiv:1911.09678

\bibitem[{{Greggio} \& {Renzini}(1983)}]{Greggio83}
{Greggio}, L., \& {Renzini}, A. 1983, \aap, 118, 217

\bibitem[{{Hamann} \& {Ferland}(1992)}]{Hamann92}
{Hamann}, F., \& {Ferland}, G. 1992, \apjl, 391, L53

\bibitem[{{Hamann} \& {Ferland}(1993)}]{Hamann93}
---. 1993, \apj, 418, 11

\bibitem[{{Hamann} \& {Ferland}(1999)}]{Hamann99}
---. 1999, \araa, 37, 487

\bibitem[{{Hamann} {et~al.}(2002){Hamann}, {Korista}, {Ferland}, {Warner}, \&
  {Baldwin}}]{Hamann02}
{Hamann}, F., {Korista}, K.~T., {Ferland}, G.~J., {Warner}, C., \& {Baldwin},
  J. 2002, \apj, 564, 592

\bibitem[{{Inayoshi} {et~al.}(2019){Inayoshi}, {Visbal}, \&
  {Haiman}}]{Inayoshi20}
{Inayoshi}, K., {Visbal}, E., \& {Haiman}, Z. 2019, arXiv e-prints,
  arXiv:1911.05791

\bibitem[{{Iwamoto} {et~al.}(1999){Iwamoto}, {Brachwitz}, {Nomoto},
  {Kishimoto}, {Umeda}, {Hix}, \& {Thielemann}}]{Iwamoto99}
{Iwamoto}, K., {Brachwitz}, F., {Nomoto}, K., {et~al.} 1999, \apjs, 125, 439

\bibitem[{{Iwamuro} {et~al.}(2004){Iwamuro}, {Kimura}, {Eto}, {Maihara},
  {Motohara}, {Yoshii}, \& {Doi}}]{Iwamuro04}
{Iwamuro}, F., {Kimura}, M., {Eto}, S., {et~al.} 2004, \apj, 614, 69

\bibitem[{{Iwamuro} {et~al.}(2002){Iwamuro}, {Motohara}, {Maihara}, {Kimura},
  {Yoshii}, \& {Doi}}]{Iwamuro02}
{Iwamuro}, F., {Motohara}, K., {Maihara}, T., {et~al.} 2002, \apj, 565, 63

\bibitem[{{Izumi} {et~al.}(2018){Izumi}, {Onoue}, {Shirakata}, {Nagao},
  {Kohno}, {Matsuoka}, {Imanishi}, {Strauss}, {Kashikawa}, {Schulze},
  {Silverman}, {Fujimoto}, {Harikane}, {Toba}, {Umehata}, {Nakanishi},
  {Greene}, {Tamura}, {Taniguchi}, {Yamaguchi}, {Goto}, {Hashimoto},
  {Ikarashi}, {Iono}, {Iwasawa}, {Lee}, {Makiya}, {Minezaki}, \&
  {Tang}}]{Izumi18}
{Izumi}, T., {Onoue}, M., {Shirakata}, H., {et~al.} 2018, \pasj, 70, 36

\bibitem[{{Izumi} {et~al.}(2019){Izumi}, {Onoue}, {Matsuoka}, {Nagao},
  {Strauss}, {Imanishi}, {Kashikawa}, {Fujimoto}, {Kohno}, {Toba}, {Umehata},
  {Goto}, {Ueda}, {Shirakata}, {Silverman}, {Greene}, {Harikane}, {Hashimoto},
  {Ikarashi}, {Iono}, {Iwasawa}, {Lee}, {Minezaki}, {Nakanishi}, {Tamura},
  {Tang}, \& {Taniguchi}}]{Izumi19}
{Izumi}, T., {Onoue}, M., {Matsuoka}, Y., {et~al.} 2019, \pasj, 71, 111

\bibitem[{{Jiang} {et~al.}(2007){Jiang}, {Fan}, {Vestergaard}, {Kurk},
  {Walter}, {Kelly}, \& {Strauss}}]{Jiang07}
{Jiang}, L., {Fan}, X., {Vestergaard}, M., {et~al.} 2007, \aj, 134, 1150

\bibitem[{{Jiang} {et~al.}(2016){Jiang}, {McGreer}, {Fan}, {Strauss},
  {Ba{\~n}ados}, {Becker}, {Bian}, {Farnsworth}, {Shen}, {Wang}, {Wang},
  {Wang}, {White}, {Wu}, {Wu}, {Yang}, \& {Yang}}]{Jiang16}
{Jiang}, L., {McGreer}, I.~D., {Fan}, X., {et~al.} 2016, \apj, 833, 222

\bibitem[{{Juarez} {et~al.}(2009){Juarez}, {Maiolino}, {Mujica}, {Pedani},
  {Marinoni}, {Nagao}, {Marconi}, \& {Oliva}}]{Juarez09}
{Juarez}, Y., {Maiolino}, R., {Mujica}, R., {et~al.} 2009, \aap, 494, L25

\bibitem[{{Kawara} {et~al.}(1996){Kawara}, {Murayama}, {Taniguchi}, \&
  {Arimoto}}]{Kawara96}
{Kawara}, K., {Murayama}, T., {Taniguchi}, Y., \& {Arimoto}, N. 1996, \apjl,
  470, L85

\bibitem[{{Kelson}(2003)}]{Kelson03}
{Kelson}, D.~D. 2003, \pasp, 115, 688

\bibitem[{{Kurk} {et~al.}(2007){Kurk}, {Walter}, {Fan}, {Jiang}, {Riechers},
  {Rix}, {Pentericci}, {Strauss}, {Carilli}, \& {Wagner}}]{Kurk07}
{Kurk}, J.~D., {Walter}, F., {Fan}, X., {et~al.} 2007, \apj, 669, 32

\bibitem[{{Maiolino} {et~al.}(2003){Maiolino}, {Juarez}, {Mujica}, {Nagar}, \&
  {Oliva}}]{Maiolino03}
{Maiolino}, R., {Juarez}, Y., {Mujica}, R., {Nagar}, N.~M., \& {Oliva}, E.
  2003, \apjl, 596, L155

\bibitem[{{Maiolino} \& {Mannucci}(2019)}]{Maiolino19}
{Maiolino}, R., \& {Mannucci}, F. 2019, \aapr, 27, 3

\bibitem[{{Maoz} \& {Mannucci}(2012)}]{Maoz12}
{Maoz}, D., \& {Mannucci}, F. 2012, \pasa, 29, 447

\bibitem[{{Maoz} {et~al.}(2014){Maoz}, {Mannucci}, \& {Nelemans}}]{Maoz14}
{Maoz}, D., {Mannucci}, F., \& {Nelemans}, G. 2014, \araa, 52, 107

\bibitem[{{Matsuoka} {et~al.}(2009){Matsuoka}, {Nagao}, {Maiolino}, {Marconi},
  \& {Taniguchi}}]{MatsuokaK09}
{Matsuoka}, K., {Nagao}, T., {Maiolino}, R., {Marconi}, A., \& {Taniguchi}, Y.
  2009, \aap, 503, 721

\bibitem[{{Matsuoka} {et~al.}(2011{\natexlab{a}}){Matsuoka}, {Nagao},
  {Maiolino}, {Marconi}, \& {Taniguchi}}]{MatsuokaK11}
---. 2011{\natexlab{a}}, \aap, 532, L10

\bibitem[{{Matsuoka} {et~al.}(2011{\natexlab{b}}){Matsuoka}, {Nagao},
  {Marconi}, {Maiolino}, \& {Taniguchi}}]{Matsuoka11}
{Matsuoka}, K., {Nagao}, T., {Marconi}, A., {Maiolino}, R., \& {Taniguchi}, Y.
  2011{\natexlab{b}}, \aap, 527, A100

\bibitem[{{Matsuoka} {et~al.}(2016){Matsuoka}, {Onoue}, {Kashikawa}, {Iwasawa},
  {Strauss}, {Nagao}, {Imanishi}, {Niida}, {Toba}, {Akiyama}, {Asami}, {Bosch},
  {Foucaud}, {Furusawa}, {Goto}, {Gunn}, {Harikane}, {Ikeda}, {Kawaguchi},
  {Kikuta}, {Komiyama}, {Lupton}, {Minezaki}, {Miyazaki}, {Morokuma},
  {Murayama}, {Nishizawa}, {Ono}, {Ouchi}, {Price}, {Sameshima}, {Silverman},
  {Sugiyama}, {Tait}, {Takada}, {Takata}, {Tanaka}, {Tang}, \&
  {Utsumi}}]{Matsuoka16}
{Matsuoka}, Y., {Onoue}, M., {Kashikawa}, N., {et~al.} 2016, \apj, 828, 26

\bibitem[{{Matsuoka} {et~al.}(2019{\natexlab{a}}){Matsuoka}, {Onoue},
  {Kashikawa}, {Strauss}, {Iwasawa}, {Lee}, {Imanishi}, {Nagao}, {Akiyama},
  {Asami}, {Bosch}, {Furusawa}, {Goto}, {Gunn}, {Harikane}, {Ikeda}, {Izumi},
  {Kawaguchi}, {Kato}, {Kikuta}, {Kohno}, {Komiyama}, {Koyama}, {Lupton},
  {Minezaki}, {Miyazaki}, {Murayama}, {Niida}, {Nishizawa}, {Noboriguchi},
  {Oguri}, {Ono}, {Ouchi}, {Price}, {Sameshima}, {Schulze}, {Shirakata},
  {Silverman}, {Sugiyama}, {Tait}, {Takada}, {Takata}, {Tanaka}, {Tang},
  {Toba}, {Utsumi}, {Wang}, \& {Yamashita}}]{Matsuoka19a}
---. 2019{\natexlab{a}}, \apjl, 872, L2

\bibitem[{{Matsuoka} {et~al.}(2019{\natexlab{b}}){Matsuoka}, {Iwasawa},
  {Onoue}, {Kashikawa}, {Strauss}, {Lee}, {Imanishi}, {Nagao}, {Akiyama},
  {Asami}, {Bosch}, {Furusawa}, {Goto}, {Gunn}, {Harikane}, {Ikeda}, {Izumi},
  {Kawaguchi}, {Kato}, {Kikuta}, {Kohno}, {Komiyama}, {Koyama}, {Lupton},
  {Minezaki}, {Miyazaki}, {Murayama}, {Niida}, {Nishizawa}, {Noboriguchi},
  {Oguri}, {Ono}, {Ouchi}, {Price}, {Sameshima}, {Schulze}, {Silverman},
  {Sugiyama}, {Tait}, {Takada}, {Takata}, {Tanaka}, {Tang}, {Toba}, {Utsumi},
  {Wang}, \& {Yamashita}}]{Matsuoka19b}
{Matsuoka}, Y., {Iwasawa}, K., {Onoue}, M., {et~al.} 2019{\natexlab{b}}, \apj,
  883, 183

\bibitem[{{Matteucci} \& {Greggio}(1986)}]{Matteucci86}
{Matteucci}, F., \& {Greggio}, L. 1986, \aap, 154, 279

\bibitem[{{Mazzucchelli} {et~al.}(2017){Mazzucchelli}, {Ba{\~n}ados},
  {Venemans}, {Decarli}, {Farina}, {Walter}, {Eilers}, {Rix}, {Simcoe},
  {Stern}, {Fan}, {Schlafly}, {De Rosa}, {Hennawi}, {Chambers}, {Greiner},
  {Burgett}, {Draper}, {Kaiser}, {Kudritzki}, {Magnier}, {Metcalfe}, {Waters},
  \& {Wainscoat}}]{Mazzucchelli17}
{Mazzucchelli}, C., {Ba{\~n}ados}, E., {Venemans}, B.~P., {et~al.} 2017, \apj,
  849, 91

\bibitem[{{Meyer} {et~al.}(2019){Meyer}, {Bosman}, \& {Ellis}}]{Meyer19}
{Meyer}, R.~A., {Bosman}, S. E.~I., \& {Ellis}, R.~S. 2019, \mnras, 487, 3305

\bibitem[{{Mortlock} {et~al.}(2011){Mortlock}, {Warren}, {Venemans}, {Patel},
  {Hewett}, {McMahon}, {Simpson}, {Theuns}, {Gonz{\'a}les-Solares}, {Adamson},
  {Dye}, {Hambly}, {Hirst}, {Irwin}, {Kuiper}, {Lawrence}, \&
  {R{\"o}ttgering}}]{Mortlock11}
{Mortlock}, D.~J., {Warren}, S.~J., {Venemans}, B.~P., {et~al.} 2011, \nat,
  474, 616

\bibitem[{{Nagao} {et~al.}(2006{\natexlab{a}}){Nagao}, {Maiolino}, \&
  {Marconi}}]{Nagao06a}
{Nagao}, T., {Maiolino}, R., \& {Marconi}, A. 2006{\natexlab{a}}, \aap, 447,
  863

\bibitem[{{Nagao} {et~al.}(2006{\natexlab{b}}){Nagao}, {Marconi}, \&
  {Maiolino}}]{Nagao06b}
{Nagao}, T., {Marconi}, A., \& {Maiolino}, R. 2006{\natexlab{b}}, \aap, 447,
  157

\bibitem[{{Nomoto} {et~al.}(2013){Nomoto}, {Kobayashi}, \&
  {Tominaga}}]{Nomoto13}
{Nomoto}, K., {Kobayashi}, C., \& {Tominaga}, N. 2013, \araa, 51, 457

\bibitem[{{Novak} {et~al.}(2019){Novak}, {Ba{\~n}ados}, {Decarli}, {Walter},
  {Venemans}, {Neeleman}, {Farina}, {Mazzucchelli}, {Carilli}, {Fan}, {Rix}, \&
  {Wang}}]{Novak19}
{Novak}, M., {Ba{\~n}ados}, E., {Decarli}, R., {et~al.} 2019, \apj, 881, 63

\bibitem[{{Onoue} {et~al.}(2019){Onoue}, {Kashikawa}, {Matsuoka}, {Kato},
  {Izumi}, {Nagao}, {Strauss}, {Harikane}, {Imanishi}, {Ito}, {Iwasawa},
  {Kawaguchi}, {Lee}, {Noboriguchi}, {Suh}, {Tanaka}, \& {Toba}}]{Onoue19}
{Onoue}, M., {Kashikawa}, N., {Matsuoka}, Y., {et~al.} 2019, \apj, 880, 77

\bibitem[{{Planck Collaboration} {et~al.}(2016){Planck Collaboration}, {Adam},
  {Aghanim}, {Ashdown}, {Aumont}, {Baccigalupi}, {Ballardini}, {Banday},
  {Barreiro}, {Bartolo}, {Basak}, {Battye}, {Benabed}, {Bernard}, {Bersanelli},
  {Bielewicz}, {Bock}, {Bonaldi}, {Bonavera}, {Bond}, {Borrill}, {Bouchet},
  {Boulanger}, {Bucher}, {Burigana}, {Calabrese}, {Cardoso}, {Carron},
  {Chiang}, {Colombo}, {Combet}, {Comis}, {Couchot}, {Coulais}, {Crill},
  {Curto}, {Cuttaia}, {Davis}, {de Bernardis}, {de Rosa}, {de Zotti},
  {Delabrouille}, {Di Valentino}, {Dickinson}, {Diego}, {Dor{\'e}}, {Douspis},
  {Ducout}, {Dupac}, {Elsner}, {En{\ss}lin}, {Eriksen}, {Falgarone}, {Fantaye},
  {Finelli}, {Forastieri}, {Frailis}, {Fraisse}, {Franceschi}, {Frolov},
  {Galeotta}, {Galli}, {Ganga}, {G{\'e}nova-Santos}, {Gerbino}, {Ghosh},
  {Gonz{\'a}lez-Nuevo}, {G{\'o}rski}, {Gruppuso}, {Gudmundsson}, {Hansen},
  {Helou}, {Henrot-Versill{\'e}}, {Herranz}, {Hivon}, {Huang}, {Ili{\'c}},
  {Jaffe}, {Jones}, {Keih{\"a}nen}, {Keskitalo}, {Kisner}, {Knox},
  {Krachmalnicoff}, {Kunz}, {Kurki-Suonio}, {Lagache}, {L{\"a}hteenm{\"a}ki},
  {Lamarre}, {Langer}, {Lasenby}, {Lattanzi}, {Lawrence}, {Le Jeune},
  {Levrier}, {Lewis}, {Liguori}, {Lilje}, {L{\'o}pez-Caniego}, {Ma},
  {Mac{\'\i}as-P{\'e}rez}, {Maggio}, {Mangilli}, {Maris}, {Martin},
  {Mart{\'\i}nez-Gonz{\'a}lez}, {Matarrese}, {Mauri}, {McEwen}, {Meinhold},
  {Melchiorri}, {Mennella}, {Migliaccio}, {Miville-Desch{\^e}nes}, {Molinari},
  {Moneti}, {Montier}, {Morgante}, {Moss}, {Naselsky}, {Natoli}, {Oxborrow},
  {Pagano}, {Paoletti}, {Partridge}, {Patanchon}, {Patrizii}, {Perdereau},
  {Perotto}, {Pettorino}, {Piacentini}, {Plaszczynski}, {Polastri}, {Polenta},
  {Puget}, {Rachen}, {Racine}, {Reinecke}, {Remazeilles}, {Renzi}, {Rocha},
  {Rossetti}, {Roudier}, {Rubi{\~n}o-Mart{\'\i}n}, {Ruiz-Granados}, {Salvati},
  {Sandri}, {Savelainen}, {Scott}, {Sirri}, {Sunyaev}, {Suur-Uski}, {Tauber},
  {Tenti}, {Toffolatti}, {Tomasi}, {Tristram}, {Trombetti}, {Valiviita}, {Van
  Tent}, {Vielva}, {Villa}, {Vittorio}, {Wandelt}, {Wehus}, {White}, {Zacchei},
  \& {Zonca}}]{Planck16}
{Planck Collaboration}, {Adam}, R., {Aghanim}, N., {et~al.} 2016, \aap, 596,
  A108

\bibitem[{{Plotkin} {et~al.}(2015){Plotkin}, {Shemmer}, {Trakhtenbrot},
  {Anderson}, {Brandt}, {Fan}, {Gallo}, {Lira}, {Luo}, {Richards}, {Schneider},
  {Strauss}, \& {Wu}}]{Plotkin15}
{Plotkin}, R.~M., {Shemmer}, O., {Trakhtenbrot}, B., {et~al.} 2015, \apj, 805,
  123

\bibitem[{{Prochaska} {et~al.}(2020){Prochaska}, {Hennawi}, {Westfall},
  {Cooke}, {Wang}, {Hsyu}, \& {Farina}}]{Pypeit20}
{Prochaska}, J.~X., {Hennawi}, J.~F., {Westfall}, K.~B., {et~al.} 2020, arXiv
  e-prints, arXiv:2005.06505

\bibitem[{{Reed} {et~al.}(2015){Reed}, {McMahon}, {Banerji}, {Becker},
  {Gonzalez-Solares}, {Martini}, {Ostrovski}, {Rauch}, {Abbott}, {Abdalla},
  {Allam}, {Benoit-Levy}, {Bertin}, {Buckley-Geer}, {Burke}, {Carnero Rosell},
  {da Costa}, {D'Andrea}, {DePoy}, {Desai}, {Diehl}, {Doel}, {Cunha},
  {Estrada}, {Evrard}, {Fausti Neto}, {Finley}, {Fosalba}, {Frieman}, {Gruen},
  {Honscheid}, {James}, {Kent}, {Kuehn}, {Kuropatkin}, {Lahav}, {Maia},
  {Makler}, {Marshall}, {Merritt}, {Miquel}, {Mohr}, {Nord}, {Ogando},
  {Plazas}, {Romer}, {Roodman}, {Rykoff}, {Sako}, {Sanchez}, {Santiago},
  {Schubnell}, {Sevilla}, {Smith}, {Soares-Santos}, {Suchyta}, {Swanson},
  {Tarle}, {Thomas}, {Tucker}, {Walker}, \& {Wechsler}}]{Reed15}
{Reed}, S.~L., {McMahon}, R.~G., {Banerji}, M., {et~al.} 2015, \mnras, 454,
  3952

\bibitem[{{Richards} {et~al.}(2006){Richards}, {Lacy}, {Storrie-Lombardi},
  {Hall}, {Gallagher}, {Hines}, {Fan}, {Papovich}, {Vanden Berk}, {Trammell},
  {Schneider}, {Vestergaard}, {York}, {Jester}, {Anderson}, {Budav{\'a}ri}, \&
  {Szalay}}]{Richards06a}
{Richards}, G.~T., {Lacy}, M., {Storrie-Lombardi}, L.~J., {et~al.} 2006, The
  Astrophysical Journal Supplement Series, 166, 470

\bibitem[{{Richards} {et~al.}(2011){Richards}, {Kruczek}, {Gallagher}, {Hall},
  {Hewett}, {Leighly}, {Deo}, {Kratzer}, \& {Shen}}]{Richards11}
{Richards}, G.~T., {Kruczek}, N.~E., {Gallagher}, S.~C., {et~al.} 2011, \aj,
  141, 167

\bibitem[{{Rodney} {et~al.}(2014){Rodney}, {Riess}, {Strolger}, {Dahlen},
  {Graur}, {Casertano}, {Dickinson}, {Ferguson}, {Garnavich}, {Hayden}, {Jha},
  {Jones}, {Kirshner}, {Koekemoer}, {McCully}, {Mobasher}, {Patel}, {Weiner},
  {Cenko}, {Clubb}, {Cooper}, {Filippenko}, {Frederiksen}, {Hjorth},
  {Leibundgut}, {Matheson}, {Nayyeri}, {Penner}, {Trump}, {Silverman}, {U},
  {Azalee Bostroem}, {Challis}, {Rajan}, {Wolff}, {Faber}, {Grogin}, \&
  {Kocevski}}]{Rodney14}
{Rodney}, S.~A., {Riess}, A.~G., {Strolger}, L.-G., {et~al.} 2014, \aj, 148, 13

\bibitem[{{Rousselot} {et~al.}(2000){Rousselot}, {Lidman}, {Cuby}, {Moreels},
  \& {Monnet}}]{Rousselot00}
{Rousselot}, P., {Lidman}, C., {Cuby}, J.~G., {Moreels}, G., \& {Monnet}, G.
  2000, \aap, 354, 1134

\bibitem[{{Sameshima} {et~al.}(2017){Sameshima}, {Yoshii}, \&
  {Kawara}}]{Sameshima17}
{Sameshima}, H., {Yoshii}, Y., \& {Kawara}, K. 2017, \apj, 834, 203

\bibitem[{{Selsing} {et~al.}(2016){Selsing}, {Fynbo}, {Christensen}, \&
  {Krogager}}]{Selsing16}
{Selsing}, J., {Fynbo}, J.~P.~U., {Christensen}, L., \& {Krogager}, J.~K. 2016,
  \aap, 585, A87

\bibitem[{{Shen}(2013)}]{Shen13_review}
{Shen}, Y. 2013, Bulletin of the Astronomical Society of India, 41, 61

\bibitem[{{Shen} {et~al.}(2016){Shen}, {Brandt}, {Richards}, {Denney},
  {Greene}, {Grier}, {Ho}, {Peterson}, {Petitjean}, {Schneider}, {Tao}, \&
  {Trump}}]{Shen16}
{Shen}, Y., {Brandt}, W.~N., {Richards}, G.~T., {et~al.} 2016, \apj, 831, 7

\bibitem[{{Shen} {et~al.}(2019){Shen}, {Wu}, {Jiang}, {Ba{\~n}ados}, {Fan},
  {Ho}, {Riechers}, {Strauss}, {Venemans}, {Vestergaard}, {Walter}, {Wang},
  {Willott}, {Wu}, \& {Yang}}]{Shen19}
{Shen}, Y., {Wu}, J., {Jiang}, L., {et~al.} 2019, \apj, 873, 35

\bibitem[{{Shin} {et~al.}(2019){Shin}, {Nagao}, {Woo}, \& {Le}}]{Shin19}
{Shin}, J., {Nagao}, T., {Woo}, J.-H., \& {Le}, H. A.~N. 2019, \apj, 874, 22

\bibitem[{{Storchi Bergmann} {et~al.}(1990){Storchi Bergmann}, {Bica}, \&
  {Pastoriza}}]{StorchiBergmann90}
{Storchi Bergmann}, T., {Bica}, E., \& {Pastoriza}, M.~G. 1990, \mnras, 245,
  749

\bibitem[{{Tang} {et~al.}(2019){Tang}, {Goto}, {Ohyama}, {Jin}, {Done}, {Lu},
  {Hashimoto}, {Kilerci Eser}, {Chiang}, \& {Kim}}]{Tang19}
{Tang}, J.-J., {Goto}, T., {Ohyama}, Y., {et~al.} 2019, \mnras, 484, 2575

\bibitem[{{Thompson} {et~al.}(1999){Thompson}, {Hill}, \&
  {Elston}}]{Thompson99}
{Thompson}, K.~L., {Hill}, G.~J., \& {Elston}, R. 1999, \apj, 515, 487

\bibitem[{{Tsuzuki} {et~al.}(2006){Tsuzuki}, {Kawara}, {Yoshii}, {Oyabu},
  {Tanab{\'e}}, \& {Matsuoka}}]{Tsuzuki06}
{Tsuzuki}, Y., {Kawara}, K., {Yoshii}, Y., {et~al.} 2006, \apj, 650, 57

\bibitem[{{van Dokkum}(2001)}]{vanDokkum01}
{van Dokkum}, P.~G. 2001, \pasp, 113, 1420

\bibitem[{{Vanden Berk} {et~al.}(2001){Vanden Berk}, {Richards}, {Bauer},
  {Strauss}, {Schneider}, {Heckman}, {York}, {Hall}, {Fan}, {Knapp},
  {Anderson}, {Annis}, {Bahcall}, {Bernardi}, {Briggs}, {Brinkmann}, {Brunner},
  {Burles}, {Carey}, {Castander}, {Connolly}, {Crocker}, {Csabai}, {Doi},
  {Finkbeiner}, {Friedman}, {Frieman}, {Fukugita}, {Gunn}, {Hennessy},
  {Ivezi{\'c}}, {Kent}, {Kunszt}, {Lamb}, {Leger}, {Long}, {Loveday}, {Lupton},
  {Meiksin}, {Merelli}, {Munn}, {Newberg}, {Newcomb}, {Nichol}, {Owen}, {Pier},
  {Pope}, {Rockosi}, {Schlegel}, {Siegmund}, {Smee}, {Snir}, {Stoughton},
  {Stubbs}, {SubbaRao}, {Szalay}, {Szokoly}, {Tremonti}, {Uomoto}, {Waddell},
  {Yanny}, \& {Zheng}}]{VB01}
{Vanden Berk}, D.~E., {Richards}, G.~T., {Bauer}, A., {et~al.} 2001, \aj, 122,
  549

\bibitem[{{Venemans} {et~al.}(2019){Venemans}, {Neeleman}, {Walter}, {Novak},
  {Decarli}, {Hennawi}, \& {Rix}}]{Venemans19}
{Venemans}, B.~P., {Neeleman}, M., {Walter}, F., {et~al.} 2019, \apj, 874, L30

\bibitem[{{Venemans} {et~al.}(2016){Venemans}, {Walter}, {Zschaechner},
  {Decarli}, {De Rosa}, {Findlay}, {McMahon}, \& {Sutherland}}]{Venemans16}
{Venemans}, B.~P., {Walter}, F., {Zschaechner}, L., {et~al.} 2016, \apj, 816,
  37

\bibitem[{{Venemans} {et~al.}(2013){Venemans}, {Findlay}, {Sutherland}, {De
  Rosa}, {McMahon}, {Simcoe}, {Gonz{\'a}lez-Solares}, {Kuijken}, \&
  {Lewis}}]{Venemans13}
{Venemans}, B.~P., {Findlay}, J.~R., {Sutherland}, W.~J., {et~al.} 2013, \apj,
  779, 24

\bibitem[{{Venemans} {et~al.}(2017{\natexlab{a}}){Venemans}, {Walter},
  {Decarli}, {Ba{\~n}ados}, {Hodge}, {Hewett}, {McMahon}, {Mortlock}, \&
  {Simpson}}]{Venemans17a}
{Venemans}, B.~P., {Walter}, F., {Decarli}, R., {et~al.} 2017{\natexlab{a}},
  ApJ, 837, 146

\bibitem[{{Venemans} {et~al.}(2017{\natexlab{b}}){Venemans}, {Walter},
  {Decarli}, {Ba{\~n}ados}, {Carilli}, {Winters}, {Schuster}, {da Cunha},
  {Fan}, {Farina}, {Mazzucchelli}, {Rix}, \& {Weiss}}]{Venemans17}
---. 2017{\natexlab{b}}, \apjl, 851, L8

\bibitem[{{Verner} {et~al.}(1999){Verner}, {Verner}, {Korista}, {Ferguson},
  {Hamann}, \& {Ferland}}]{Verner99}
{Verner}, E.~M., {Verner}, D.~A., {Korista}, K.~T., {et~al.} 1999, \apjs, 120,
  101

\bibitem[{{Vestergaard} \& {Osmer}(2009)}]{Vestergaard09}
{Vestergaard}, M., \& {Osmer}, P.~S. 2009, \apj, 699, 800

\bibitem[{{Vestergaard} \& {Wilkes}(2001)}]{Vestergaard01}
{Vestergaard}, M., \& {Wilkes}, B.~J. 2001, The Astrophysical Journal
  Supplement Series, 134, 1

\bibitem[{{Vietri} {et~al.}(2018){Vietri}, {Piconcelli}, {Bischetti}, {Duras},
  {Martocchia}, {Bongiorno}, {Marconi}, {Zappacosta}, {Bisogni}, {Bruni},
  {Brusa}, {Comastri}, {Cresci}, {Feruglio}, {Giallongo}, {La Franca},
  {Mainieri}, {Mannucci}, {Ricci}, {Sani}, {Testa}, {Tombesi}, {Vignali}, \&
  {Fiore}}]{Vietri18}
{Vietri}, G., {Piconcelli}, E., {Bischetti}, M., {et~al.} 2018, \aap, 617, A81

\bibitem[{{Wang} {et~al.}(2018){Wang}, {Yang}, {Fan}, {Yue}, {Wu}, {Schindler},
  {Bian}, {Li}, {Farina}, {Ba{\~n}ados}, {Davies}, {Decarli}, {Green}, {Jiang},
  {Hennawi}, {Huang}, {Mazzucchelli}, {McGreer}, {Venemans}, {Walter}, \&
  {Beletsky}}]{Wang18}
{Wang}, F., {Yang}, J., {Fan}, X., {et~al.} 2018, \apj, 869, L9

\bibitem[{{Wang} {et~al.}(2013){Wang}, {Wagg}, {Carilli}, {Walter}, {Lentati},
  {Fan}, {Riechers}, {Bertoldi}, {Narayanan}, {Strauss}, {Cox}, {Omont},
  {Menten}, {Knudsen}, {Neri}, \& {Jiang}}]{Wang13}
{Wang}, R., {Wagg}, J., {Carilli}, C.~L., {et~al.} 2013, ApJ, 773, 44

\bibitem[{{Wang} {et~al.}(2016){Wang}, {Wu}, {Neri}, {Fan}, {Walter},
  {Carilli}, {Momjian}, {Bertoldi}, {Strauss}, {Li}, {Wang}, {Riechers},
  {Jiang}, {Omont}, {Wagg}, \& {Cox}}]{Wang16}
{Wang}, R., {Wu}, X.-B., {Neri}, R., {et~al.} 2016, \apj, 830, 53

\bibitem[{{Willott} {et~al.}(2010){Willott}, {Delorme}, {Reyl{\'e}}, {Albert},
  {Bergeron}, {Crampton}, {Delfosse}, {Forveille}, {Hutchings}, {McLure},
  {Omont}, \& {Schade}}]{Willott10b}
{Willott}, C.~J., {Delorme}, P., {Reyl{\'e}}, C., {et~al.} 2010, \aj, 139, 906

\bibitem[{{Woo} {et~al.}(2018){Woo}, {Le}, {Karouzos}, {Park}, {Park},
  {Malkan}, {Treu}, \& {Bennert}}]{Woo18}
{Woo}, J.-H., {Le}, H. A.~N., {Karouzos}, M., {et~al.} 2018, \apj, 859, 138

\bibitem[{{Wu} {et~al.}(2015){Wu}, {Wang}, {Fan}, {Yi}, {Zuo}, {Bian}, {Jiang},
  {McGreer}, {Wang}, {Yang}, {Yang}, {Thompson}, \& {Beletsky}}]{Wu15}
{Wu}, X.-B., {Wang}, F., {Fan}, X., {et~al.} 2015, \nat, 518, 512

\bibitem[{{Xu} {et~al.}(2018){Xu}, {Bian}, {Shen}, {Zuo}, {Fan}, \&
  {Zhu}}]{Xu18}
{Xu}, F., {Bian}, F., {Shen}, Y., {et~al.} 2018, \mnras, 480, 345

\bibitem[{{Yang} {et~al.}(2019){Yang}, {Wang}, {Fan}, {Yue}, {Wu}, {Li},
  {Bian}, {Jiang}, {Ba{\~n}ados}, \& {Beletsky}}]{Yang19}
{Yang}, J., {Wang}, F., {Fan}, X., {et~al.} 2019, \aj, 157, 236

\bibitem[{{York} {et~al.}(2000){York}, {Adelman}, {Anderson}, {Anderson},
  {Annis}, {Bahcall}, {Bakken}, {Barkhouser}, {Bastian}, {Berman}, {Boroski},
  {Bracker}, {Briegel}, {Briggs}, {Brinkmann}, {Brunner}, {Burles}, {Carey},
  {Carr}, {Castander}, {Chen}, {Colestock}, {Connolly}, {Crocker}, {Csabai},
  {Czarapata}, {Davis}, {Doi}, {Dombeck}, {Eisenstein}, {Ellman}, {Elms},
  {Evans}, {Fan}, {Federwitz}, {Fiscelli}, {Friedman}, {Frieman}, {Fukugita},
  {Gillespie}, {Gunn}, {Gurbani}, {de Haas}, {Haldeman}, {Harris}, {Hayes},
  {Heckman}, {Hennessy}, {Hindsley}, {Holm}, {Holmgren}, {Huang}, {Hull},
  {Husby}, {Ichikawa}, {Ichikawa}, {Ivezi{\'c}}, {Kent}, {Kim}, {Kinney},
  {Klaene}, {Kleinman}, {Kleinman}, {Knapp}, {Korienek}, {Kron}, {Kunszt},
  {Lamb}, {Lee}, {Leger}, {Limmongkol}, {Lindenmeyer}, {Long}, {Loomis},
  {Loveday}, {Lucinio}, {Lupton}, {MacKinnon}, {Mannery}, {Mantsch}, {Margon},
  {McGehee}, {McKay}, {Meiksin}, {Merelli}, {Monet}, {Munn}, {Narayanan},
  {Nash}, {Neilsen}, {Neswold}, {Newberg}, {Nichol}, {Nicinski}, {Nonino},
  {Okada}, {Okamura}, {Ostriker}, {Owen}, {Pauls}, {Peoples}, {Peterson},
  {Petravick}, {Pier}, {Pope}, {Pordes}, {Prosapio}, {Rechenmacher}, {Quinn},
  {Richards}, {Richmond}, {Rivetta}, {Rockosi}, {Ruthmansdorfer}, {Sand ford},
  {Schlegel}, {Schneider}, {Sekiguchi}, {Sergey}, {Shimasaku}, {Siegmund},
  {Smee}, {Smith}, {Snedden}, {Stone}, {Stoughton}, {Strauss}, {Stubbs},
  {SubbaRao}, {Szalay}, {Szapudi}, {Szokoly}, {Thakar}, {Tremonti}, {Tucker},
  {Uomoto}, {Vanden Berk}, {Vogeley}, {Waddell}, {Wang}, {Watanabe},
  {Weinberg}, {Yanny}, {Yasuda}, \& {SDSS Collaboration}}]{York00}
{York}, D.~G., {Adelman}, J., {Anderson}, John~E., J., {et~al.} 2000, \aj, 120,
  1579

\bibitem[{{Yoshii} {et~al.}(1998){Yoshii}, {Tsujimoto}, \& {Kawara}}]{Yoshii98}
{Yoshii}, Y., {Tsujimoto}, T., \& {Kawara}, K. 1998, \apjl, 507, L113

\end{thebibliography}




\end{document}